\documentclass[11pt,twocolumn,reprint,superscriptaddress,aps,nofootinbib,nolongbibliography,pre]{revtex4-2}

\usepackage{url,psfrag,graphicx}
\usepackage{amssymb}
\usepackage{bm}
\usepackage{caption}
\usepackage{dcolumn}
\usepackage{hyperref}
\usepackage{float,epsfig}
\usepackage{mathtools}
\usepackage{times}
\usepackage{multirow}

\begin{document}

\title{Universal interface fluctuations in the contact process}

\author{B.\ G.\ Barreales}
\affiliation{Departamento de F\'{\i}sica, Universidad de Extremadura, 06006 Badajoz, Spain}
\author{J.\ J.\ Mel\'endez}
\affiliation{Departamento de F\'{\i}sica, Universidad de Extremadura, 06006 Badajoz, Spain}
\address{Instituto de Computaci\'on Cient\'{\i}fica Avanzada de Extremadura (ICCAEx), Universidad de Extremadura, 06006 Badajoz, Spain}
\author{R.\ Cuerno}
\affiliation{Departamento de Matem\'aticas and Grupo Interdisciplinar de Sistemas Complejos (GISC), Universidad Carlos III de Madrid, 28911 Legan\'es, Spain}
\author{J.\ J.\ Ruiz-Lorenzo}
\affiliation{Departamento de F\'{\i}sica, Universidad de Extremadura, 06006 Badajoz, Spain}
\address{Instituto de Computaci\'on Cient\'{\i}fica Avanzada de Extremadura (ICCAEx), Universidad de Extremadura, 06006 Badajoz, Spain}

\date{October 3, 2022}

\begin{abstract}
We study the interface representation of the contact process (CP) at its directed-percolation critical point, where the scaling properties of the interface can be related to those of the original particle model. Interestingly, such a behavior happens to be intrinsically anomalous and more complex than that described by the standard Family-Vicsek dynamic scaling Ansatz of surface kinetic roughening. We expand on a previous numerical study by Dickman and Muñoz [Phys.\ Rev.\ E \textbf{62}, 7632 (2000)] to fully characterize the kinetic roughening universality class for interface dimensions $d=1, 2$, and 3. Beyond obtaining scaling exponent values, we characterize the interface fluctuations via their probability density function (PDF) and covariance, seen to display universal properties which are qualitatively similar to those recently assessed for the Kardar-Parisi-Zhang (KPZ) and other important universality classes of kinetic roughening. Quantitatively, while for $d=1$ the interface covariance seems to be well described by the KPZ, Airy$_1$ covariance, no such agreement occurs in terms of the fluctuation PDF nor the scaling exponents. 

\end{abstract} 

\maketitle 
\section{Introduction}
\label{sect:introduction}

Many spatially-extended systems of a high current interest operate far from equilibrium. The conditions for and the properties of the emergence of the strong correlations associated with space-time criticality \cite{Tauber2014} become particularly relevant in that context. 
Among the various known modes of criticality far from equilibrium, surface kinetic roughening \cite{Barabasi1995,Krug1997} stands out due to its ubiquity throughout science. In principle, this phenomenon refers to the critical fluctuations of the interface of a driven system which is subject to some kind of noise. However, the ensuing universality classes and their properties are being quite recently seen to generalize and expand \cite{kriecherbauer2010,Halpin-Healy2015,Takeuchi2018} those of equilibrium critical dynamics to non-equilibrium conditions \cite{Tauber2014}, becoming relevant even for non-interfacial systems. 

For instance, as implied by recent results for the celebrated Kardar-Parisi-Zhang (KPZ) \cite{kriecherbauer2010,Halpin-Healy2015,Takeuchi2018} and others universality classes \cite{Barabasi1995,Krug1997,Carrasco2019, Carrasco2016,Carrasco2019,Rodriguez-Fernandez2020}, it is becoming increasingly clear that, beyond scaling exponent values, unambiguous characterization of kinetic roughening universality classes requires assessing also the statistics of interface fluctuations via their probability distribution function (PDF) and covariance. Once the time dependence of the interface fluctuations is suitably rescaled out, such functions happen to also be universal \cite{kriecherbauer2010,Halpin-Healy2012,Halpin-Healy2013,Oliveira2012,Oliveira2013,Halpin-Healy2014,Almeida2014,Halpin-Healy2015,Takeuchi2018}, leading to values of the cumulants of the PDF, like the skewness and kurtosis, which characterize the universality class, much like amplitude ratios do for equilibrium critical systems \cite{Henkel2008}; see Refs.\ \cite{Rodriguez-Fernandez2021,Marcos2022} for some recent discussions. 

An example of this behavior is the one-dimensional KPZ universality class, where the fluctuation PDF belongs to the well-known Tracy-Widom family of distributions for the largest eigenvalues of Hermitian random matrices \cite{Fortin2015} and the height covariance follows the Airy covariances \cite{kriecherbauer2010,Halpin-Healy2015,Takeuchi2018}. These properties are believed to generalize to all interface dimensions $d$ \cite{Halpin-Healy2012,Halpin-Healy2013,Oliveira2013,Alves2014}.

Kinetic roughening is also quite innovative with respect to the types of dynamic scaling Ans\"atze that can occur. The Family-Vicsek (FV) Ansatz \cite{Barabasi1995,Krug1997} analogous to the critical dynamics of the Ising model \cite{Tauber2014}, accurately describes the behavior of the KPZ and many other universality classes \cite{Barabasi1995,Krug1997}. 
However, generalizations of the FV Ansatz, collectively referred to as anomalous kinetic roughening, are indeed possible \cite{Schroeder93,Dassarma94,Lopez1997b,Ramasco2000}, and actually required to account for the properties of still many other (including experimental) kinetically rough systems \cite{Cuerno2004}, a very recent example being the so-called tensionless KPZ equation \cite{Rodriguez-Fernandez2022}. 

While interface fluctuation statistics have been extensively studied for universality classes that satisfy the Family-Vicsek (FV) scaling, there is a lack of research on systems that exhibit anomalous scaling asymptotically. To the best of our knowledge, there are relatively few studies addressing fluctuation statistics in such systems \cite{Rodriguez-Fernandez2022,Marcos2022,Gutierrez2023}, and none of them specifically investigate the $d$-dependent behavior of these fluctuations. It is important to note that the so-called intrinsic anomalous scaling has been argued (based on perturbative arguments) not to be asymptotic for systems with local interactions in absence of morphological instabilities and/or quenched noise \cite{Lopez2005}, hence some additional conditions are expected for it to occur.

A very interesting particle model in this context is the contact process (CP). Having been introduced to describe epidemic spreading without immunization \cite{Liggett1985}, the contact process happens to host a phase transition to an absorbing state, with the transition being in the nonequilibrium universality class of directed percolation (DP) \cite{Odor2004,Henkel2008}. Due to the existence of analytical results for this model (albeit in absence of an exact solution for it), it is the chosen realization of DP for a large research community. 

As it turns out, a fruitful direct mapping can be established between the CP and an interface model (see, e.g.,\ Ref.\ \cite{Dickman2000} and others therein), in such a way that the absorbing state of the particle model ---which corresponds to the global absence of activity--- corresponds to arrested overall motion of the corresponding interface (pinning). Right at criticality, the scaling properties of the interface can be related to those of the original particle model, thus providing information on interface dynamics at an absorbing-state critical point. 

This connection has been exploited by Dickman and Mu\~noz \cite{Dickman2000} to investigate the ensuing kinetic roughening properties of the CP for interface dimensions $d=1, 2,$ and 3 where nontrivial scaling is expected, i.e.\ below the upper critical dimension, $d_c=4$. Interestingly, the result of Ref.\ \cite{Dickman2000} is that intrinsic anomalous scaling occurs for all these values of $d$, and moreover that (some of) the kinetic roughening exponents are directly given by those describing the decay of the order parameter at the phase transition point.

In this paper we revisit the work by Dickman and Mu\~noz \cite{Dickman2000} with several aims: {\it (i)} to verify, on a paradigmatic model related with DP, if the PDF and covariance of interface fluctuations remain universal for all $d<d_c$, as is the case e.g.\ for KPZ, in spite of the dynamic scaling Ansatz not being FV; note in passing that the behavior of intrinsic anomalous scaling with interface dimension has been very scarcely assessed in the literature (see Refs.\ \cite{Szendro2007,Song2021} and other therein for some examples in which fluctuation statistics were not characterized). {\it (ii)} To accomplish the previous objective, we thoroughly assess the dynamic scaling Ansatz and scaling exponents reported in Ref.\ \cite{Dickman2000}, and provide more explicit data on the behavior of the various observables studied, in particular for $d\ge 2$; and {\it (iii)} to assess possible connections between the thus determined fluctuation statistics with those of important reference cases like the 1D KPZ universality class. 
We will find both similarities and differences. At this, note also that DP and KPZ are two paradigmatic universality classes for nonequilibrium systems \cite{Odor2004,Henkel2008} which feature a subtle interplay, as exemplified by the depinning transition of the KPZ equation with quenched disorder (see \ Refs.\ \cite{Barabasi1995,Wiese2022,Barreales2022} and others therein).

For completeness, the results reported herein will be compared with available experimental data. Although there is surprisingly little experimental confirmation of predicted universal characteristics of DP criticality, some excellent works \cite{Rupp2003,Takeuchi2007,Takeuchi2009,Lemoult2016} do report unambiguous observations of it. These references provide substantial experimental support for the connection between theoretical or numerical simulations and experiments, reinforcing the potential relevance and applicability of our findings.

The paper is organized as follows. In Sec.\ \ref{sect:model} we recall the definition of the CP as a particle model and its mapping to an interface model, together with those facts about the phase transition which are of direct relevance to the present work. Section \ref{sect:observables} collects the definitions of the observables that will be employed here, together with the most salient features of surface kinetic roughening required to rationalize our numerical data. The results from our simulations are reported in Sec.\ \ref{sect:results}, which is followed by a discussion in Sec.\ \ref{sect:disc}. Finally, Sec.\ \ref{sect:summ} contains a summary of our results, together with our conclusions. Further additional details on and results from our simulations are collected in three appendices.

\section{Model}
\label{sect:model}
The contact process (CP) is  originally defined as a particle model \cite{Liggett1985,Odor2004,Henkel2008}. Each site of the $d$-dimensional integer lattice $\mathbb{Z}^d$ is either occupied by a particle or empty. Particles are created at vacant sites at a rate which is proportional to the number of occupied nearest neighbours, and  are annihilated at a constant rate, which is normalized to be 1. If $\lambda$ quantifies the creation rate, there is a phase transition at a critical value $\lambda_c$ to an absorbing state which is empty \cite{Liggett1985,Odor2004,Henkel2008}. As noted above, one interpretation of this process is as a model for the spread of an infection, where the occupied sites and the empty ones are identified as infected and healthy individuals, respectively~\cite{Liggett1985,Odor2004,Henkel2008}.

The simulations have been carried out on a lattice of volume $L^d$ where $d$ is the dimension ($d=1,2$ or $3$), with periodic boundary conditions. Let $s$ be a binary variable which monitors the occupation or activity of a site of the lattice: at time $t$, each site (with coordinate $\boldsymbol{x}\in \mathbb{Z}^d$) can be occupied by a particle, so that $s(\boldsymbol{x},t)=1$, or empty so that $s(\boldsymbol{x}, t)=0$. At the start ($t=0$) all sites are occupied. At each time step, an occupied site is chosen randomly and two processes are possible: creation or annihilation of a particle. With probability $p=\lambda/(1+\lambda)$, one of the $2d$ nearest neighbors of the chosen site is selected; if it is empty, creation occurs. On the other hand, with probability $1-p$ the particle of the chosen site is annihilated. The time increment at each step is $\Delta t=1/N_{\mathrm{occ}}$ where $N_{\mathrm{occ}}$ is the number of occupied sites. 

To connect the previous particle model with an interface model, we define the front or interface at each lattice site as $h(\boldsymbol{x},t)=\int_0^{t}s(\boldsymbol{x},t')dt'$. Therefore, the front quantifies the total activity at site $\boldsymbol{x}$ up to time $t$. Indeed, in this way the absorbing state of the particle model, characterized by the global absence of activity, corresponds to an arrested or pinned interface. Note that, while pinning transitions are frequently found in surface kinetic roughening processes with quenched noise \cite{Barabasi1995,Wiese2022}, in the present model there are no explicit sources of quenched disorder. In this work, all our simulations are performed at the critical value of the coupling, namely, $\lambda_c=3.297847, 1.6488$, and $1.3169$, in one, two, and three dimensions, respectively \cite{Dickman2000}.

\section{Observables}
\label{sect:observables}
The order parameter for the phase transition in the contact process is the global particle density $\rho(t)$. At the critical point, this function decays as a power law, $\rho(t) \sim t^{-\theta}$, with an universal $d$-dependent critical exponent $\theta$. The exponent values of the directed percolation universality class governing this transition are shown in Table \ref{tab:DP} \cite{Henkel2008}.

\begin{table}[b]
\caption{Critical exponents of the DP universality class \cite{Henkel2008}, kinetic roughening exponents $\alpha$, $\beta$, and $\alpha_{\rm loc}$ as obtained in Ref.\ \cite{Dickman2000}, and maximum value of $L$ ($L_{\rm max}$) employed in the latter, for the values of $d$ considered in our work.}
\label{tab:DP}
\begin{ruledtabular}
\begin{tabular}{dcccccc}
d & $\theta$ \cite{Henkel2008} & $z$ \cite{Henkel2008} & $\alpha$ \cite{Dickman2000} & $\beta$ \cite{Dickman2000} & $\alpha_{\rm loc}$ \cite{Dickman2000} & $L_{\rm max}$ \cite{Dickman2000}\\
\colrule
1 & 0.159464(6) & 1.580745(10) & 1.33(1) & 0.839(1) & 0.63(3) & 5000 \\  
2 & 0.4505(10) & 1.7660(16) & 0.97(1) & 0.550(5) & 0.385(5) & 256 \\
3 & 0.732(4) & 1.901(5) & 0.51(1) & 0.27(1) & 0.09(2) & 50 \\
\end{tabular}
\end{ruledtabular}
\end{table}

\begin{figure}[h]
    \centering
    \includegraphics[width=0.45\textwidth]{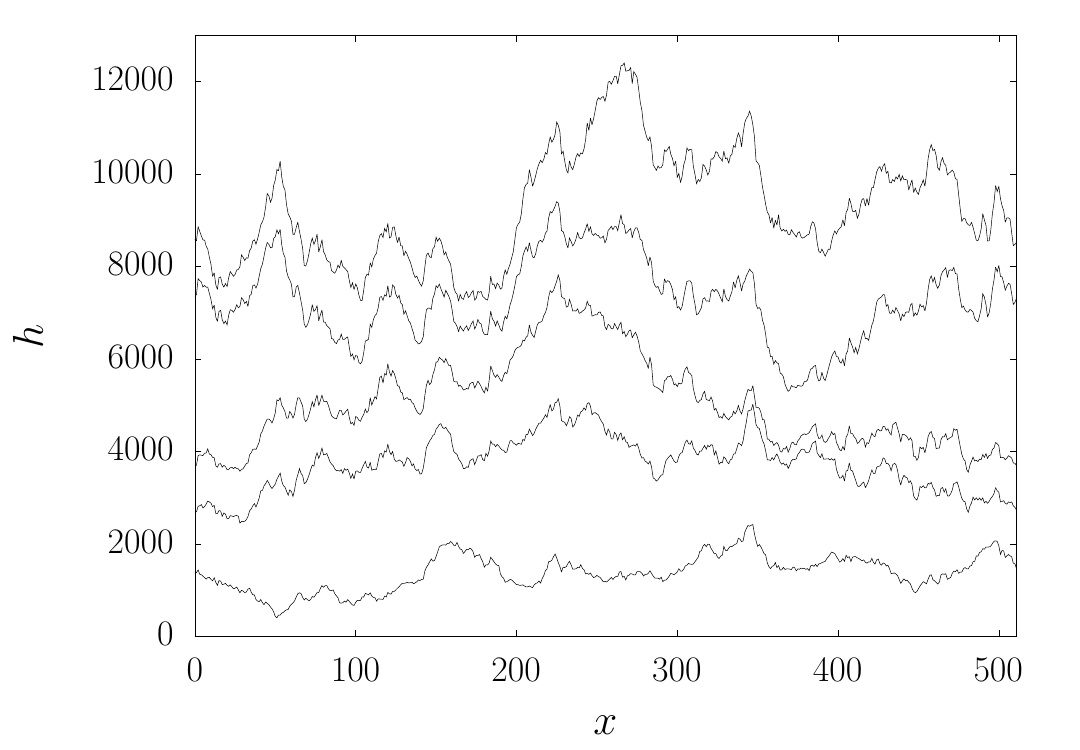}
    \caption{Local front or height profiles $h(x,t)$ from numerical simulations of a one-dimensional system with $L=512$. Each line corresponds to a different time (bottom to top): $t=5000,15000, 25000, 35000$, and $45000$. All units are arbitrary.}
    \label{fig:interface}
\end{figure}

As mentioned above, here we will rather study the associated interface problem. Thus, at each node of the lattice $\boldsymbol{x}$, the local height $h(\boldsymbol{x},t)$ is defined as the amount of time (up to the time $t$) that site has been occupied. The collection of height values $h(\boldsymbol{x},t)$ for all $\boldsymbol{x}$ defines the front at time $t$. See Fig.\ \ref{fig:interface} for sample images of fronts taken from our $d=1$ simulations. 

The mean height of the front is computed as 
\begin{equation}
    \overline{h}(t)=\frac{1}{L^d} \sum_{\boldsymbol{x}} h(\boldsymbol{x},t)\,.
\end{equation}

Fluctuations of the height field around this value can be characterized by the front width (or front roughness) $w(L,t)$, which is defined as
\begin{equation}
w^2(L,t)=\Biggl \langle \frac{1}{L^d} \sum_{\boldsymbol{x}} \left[ h(\boldsymbol{x},t)- \overline{h}(t) \right]^2 \Biggr \rangle \,,
\end{equation}
where $\langle (\cdots) \rangle$ denotes statistical average. 

For kinetically rough interfaces, the roughness is expected to satisfy the Family-Vicsek (FV) dynamic scaling relation \cite{Barabasi1995, Krug1997}
\begin{equation}
w(L,t)=t^{\beta}f(t/L^z) \,,    
\label{eq:wFV}
\end{equation}
in such a way that $w\sim t^{\beta}$ for small times such that $t\ll L^z$ and a steady-state value is achieved, $w=w_{\mathrm{sat}}\sim L^{\alpha}$, for long times such that $t\gg L^z$. 

The short- and long-time behaviors can equivalently be cast in terms of a lateral correlation length, $\xi(t)$, defined as
\begin{equation}
    \label{eq:corr_length}
    \xi(t) \sim t^{1/z},
\end{equation}
in such a way that $\xi(t)\ll L$ ($\gg L$) for short (long) times. 

The exponents $\beta$, $\alpha$, and $z$ in Eqs.\ \eqref{eq:wFV} and \eqref{eq:corr_length} are called the growth, roughness, and dynamic exponents, respectively, and are related through $\beta=\alpha/z$ \cite{Barabasi1995,Krug1997}, so that only two of them are independent. The values of these scaling exponents at the DP phase transition are known for $d<d_c$; for ease of later comparison, Table \ref{tab:DP} summarizes the values of $z$ (from properties of DP \cite{Henkel2008}), and $\alpha$ and $\beta$ as obtained in Ref.\ \cite{Dickman2000}.

\begin{figure*}[htb]
    \centering
    \includegraphics[width=0.33\textwidth]{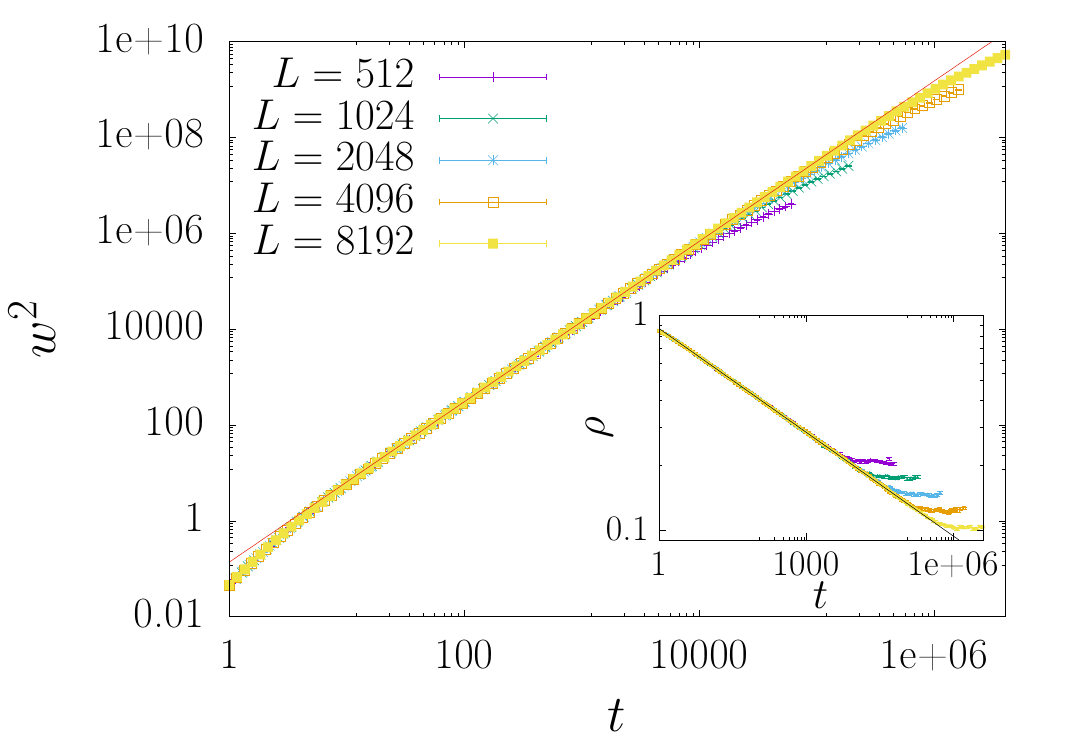}
    \includegraphics[width=0.33\textwidth]{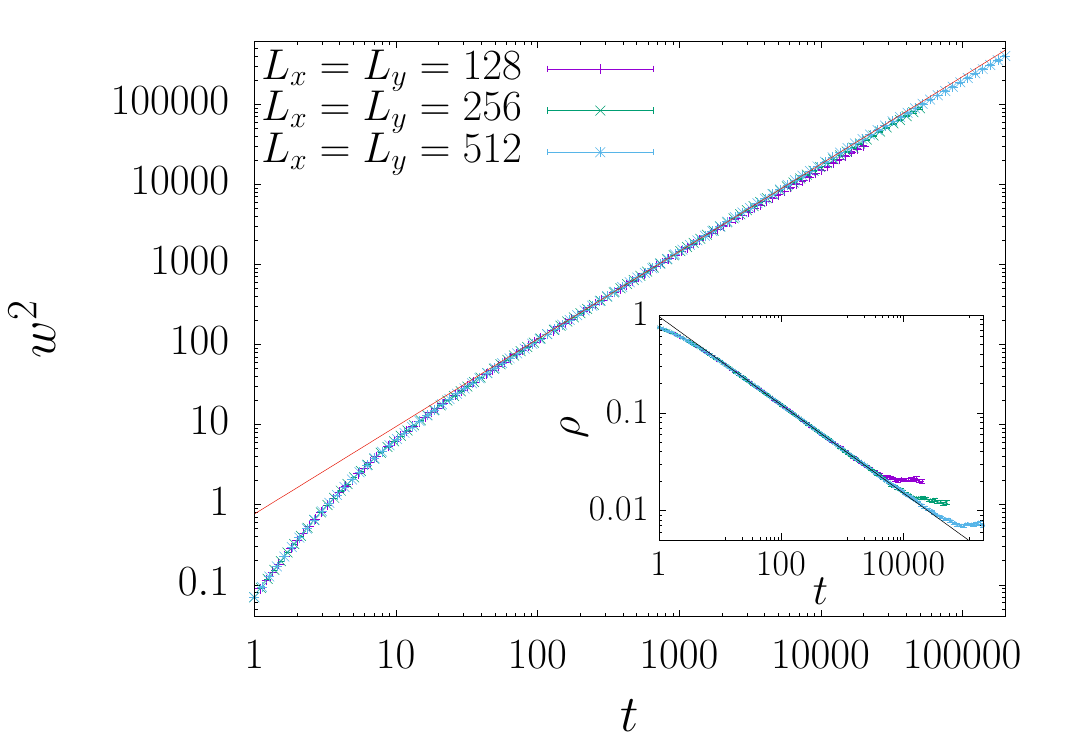}
    \includegraphics[width=0.33\textwidth]{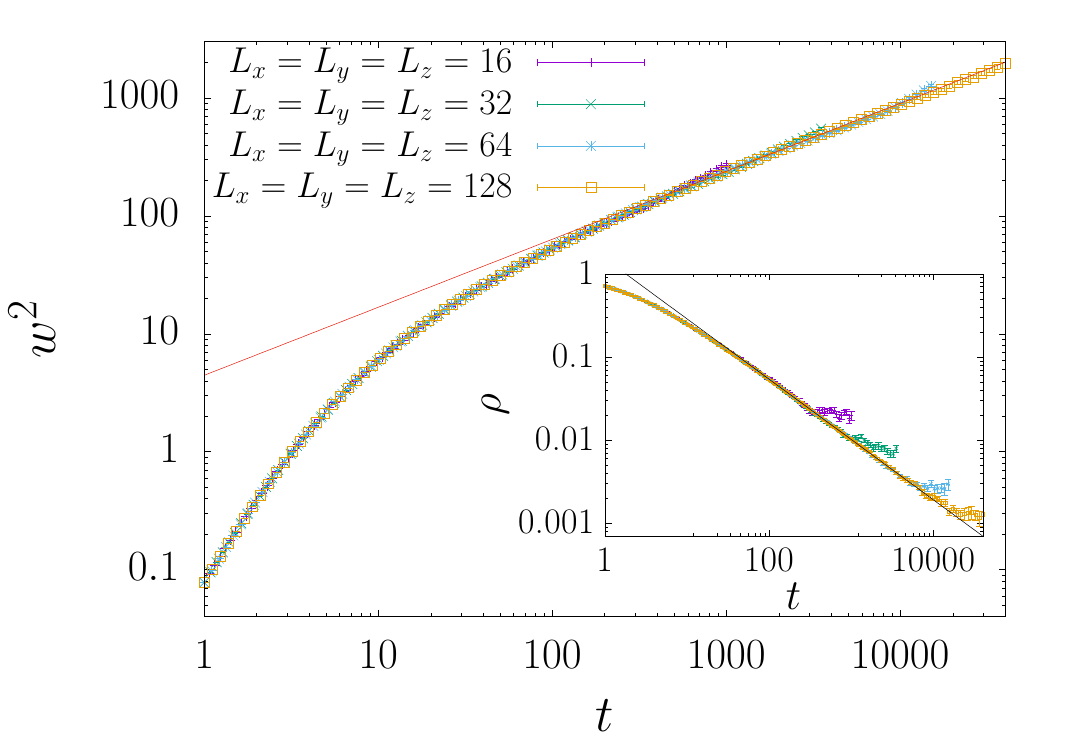}
    \caption{Squared front roughness (and particle density in the insets) versus time for $d=1, 2$, and $3$, left to right. The solid lines represent the scaling law $t^{2\beta}$ (and $t^{-\theta}$ in the insets) with the exponents corresponding to the largest sizes in Table \ref{tab:exponents}.}
    \label{fig:w2}
\end{figure*}

As in equilibrium critical dynamics \cite{Tauber2014}, for kinetic roughening systems scaling behavior also reflects into the behavior of correlation functions \cite{Barabasi1995,Krug1997}. Here, we will consider the height-difference correlation function $C_2(\boldsymbol{r},t)$, defined as
\begin{equation}
\label{eq:correlation_2}
\begin{split}
	C_2(\boldsymbol{r}, t) & =\frac{1}{L^d} \sum_{\boldsymbol{x}} \left\langle [h(\boldsymbol{x}+\boldsymbol{r}, t)
 -h(\boldsymbol{x}, t)]^2 \right\rangle.
\end{split}
\end{equation}
In two and three dimensions, we have computed these correlations for $\boldsymbol{r}$ varying in the $x$-direction only and averaging in the remaining $d-1$ directions. Under kinetic roughening conditions, the FV dynamic scaling Ansatz implies for $C_2$
\begin{equation}
    C_2(\boldsymbol{r},t)=r^{2\alpha} g_{\mathrm{FV}}(r/\xi(t)) ,
    \label{eq:corr_lengthFV}
\end{equation}
where $g_{\mathrm{FV}}$ is a scaling function which behaves as $g_{\mathrm{FV}}(u) \sim u^{-2\alpha}$ for $u\gg 1$ and $g_{\mathrm{FV}}(u)\sim {\rm const}$ for $u\ll1$ \cite{Barabasi1995,Krug1997}. In this way, for $r$ smaller than the correlation length $C_2(r,t)\sim r^{2\alpha}$ scales with distance while, for $r$ greater than the correlation length, $C_2(r,t)$ reaches a plateau $C_{2,p}(t)$ and becomes $r$-independent, so that
\begin{equation}
C_{2,p} \sim \xi^{2\alpha}   \ \ \ \ \ \mathrm{for} \ \  r \gg \xi(t)\,.
\label{eq:c2p}
\end{equation}
Moreover, one can evaluate the correlation length using 
\begin{equation}
    C_2(\xi_a(t),t)=a C_{2,p}(t),
\end{equation}
where $a$ is a constant taken arbitrarily; the precise value of which does not modify the scaling behavior \cite{Barreales2020}.

There are kinetically rough systems in which the height-difference correlation function exhibits an anomalous behavior which does not agree with the FV form given by Eq.\ \eqref{eq:corr_lengthFV}. This FV scaling  needs to be generalized into \cite{Lopez1997,Ramasco2000,Cuerno2004}
\begin{equation}
    C_2(\boldsymbol{r},t)=r^{2\alpha} g(r/\xi(t))\,,
    \label{eq:c2an}
\end{equation}
where now the new scaling function $g(u)$ behaves as $g(u) \sim u^{-2\alpha}$ for $u\gg 1$ and $g(u)\sim u^{-2(\alpha-\alpha_{\rm loc})}$ for $u\ll1$; specifically, it is not constant for small arguments, so that now $C_2(\boldsymbol{r},t)\sim r^{2\alpha_{\rm loc}}$ for small scales $r\ll \xi(t)$, from which a new exponent appears, $\alpha_{\rm loc}$, called the local roughness exponent, which characterizes the front fluctuation measured at {\em local} distances smaller than the system size $L$. For the reader's convenience, Table \ref{tab:DP} collects the values of $\alpha_{\rm loc}$ obtained for the CP in Ref.\ \cite{Dickman2000} (where this exponent is termed $\alpha_2$).

Note that the FV scaling, Eq.\ \eqref{eq:corr_lengthFV}, indeed corresponds to the particular case of Eq.\ \eqref{eq:c2an} in which $\alpha_{\rm loc}=\alpha$. When $\alpha_{\rm loc}\neq \alpha$ are independent exponents and $\alpha_{\rm loc}<1$,
scaling is said to be intrinsically anomalous \cite{Lopez1997,Cuerno2004}. Other forms of anomalous scaling are found \cite{Ramasco2000,Cuerno2004}, but they will not be of concern for our present work.

As noted in Sec.\ \ref{sect:introduction}, recent work on the KPZ \cite{kriecherbauer2010,Halpin-Healy2015,Takeuchi2018} and other kinetic roughening universality classes has shown that additional quantities also exhibit universal properties, such as the PDF of front fluctuations or the height covariance. Actually, these quantities are becoming necessary to assess the precise universality class a given system belongs to, as in some cases the values of the critical exponents may turn out to be insufficient, see Refs.\ \cite{Rodriguez-Fernandez2021,Marcos2022} for discussions and some examples. Specifically, the front fluctuations are computed as the difference between the local heights and the mean front height over different realizations. In order to achieve a universal (time-dependent) distribution for these fluctuations, one needs to normalize such a difference by the systematic increase of the fluctuations with time, namely, by the roughness. Hence, we will compute the PDF of the rescaled fluctuation variable 
\begin{equation}
\label{eq:fluc}
    \chi(\boldsymbol{x},t)=\dfrac{h(\boldsymbol{x},t)-\langle \overline{h}(t) \rangle}{t^\beta} .    
\end{equation}
as well as its skewness $s$ and excess kurtosis $k$, computed as functions of the local height fluctuation $\delta h = h(x,t)-\bar h(t)$ as $s=\langle \delta h^3 \rangle_c / \langle \delta h^2 \rangle_c^{3/2}$ and $k=\langle \delta h^4 \rangle_c / \langle \delta h^2 \rangle_c^{2}$, where $\langle \cdots \rangle_c$ denotes the cumulant average.

In addition to its PDF, the (two-point) front statistics is frequently provided \cite{Takeuchi2018} in terms of the
height covariance correlation function $C_1(\boldsymbol{r}, t)$, which is defined as
\begin{equation}
\label{eq:correlation_1}
	C_1(\boldsymbol{r}, t) =\frac 1{L^d} \sum_{\boldsymbol{x} }\left\langle h(\boldsymbol{x}+\boldsymbol{r}, t)h(\boldsymbol{x}, t) \right\rangle - \langle \overline{h}(t) \rangle^2\,.
\end{equation}
As for the case of the height-difference correlation function $C_2$, for $d>1$ we have taken $\boldsymbol{r}$ only in the $x$-direction, averaging in the remaining $d-1$ directions. Also note that the three observables defined in this section are mathematically related. Indeed, under the assumption of rotational invariance, so that dependence on $\boldsymbol{r}$ is only through $r=|\boldsymbol{r}|$, one has that $C_2(r,t)= 2 [w^2(t)-C_1(r,t)]$ \cite{Krug1997}.

\begin{figure*}[bt]
    \centering
    \includegraphics[width=0.33\textwidth]{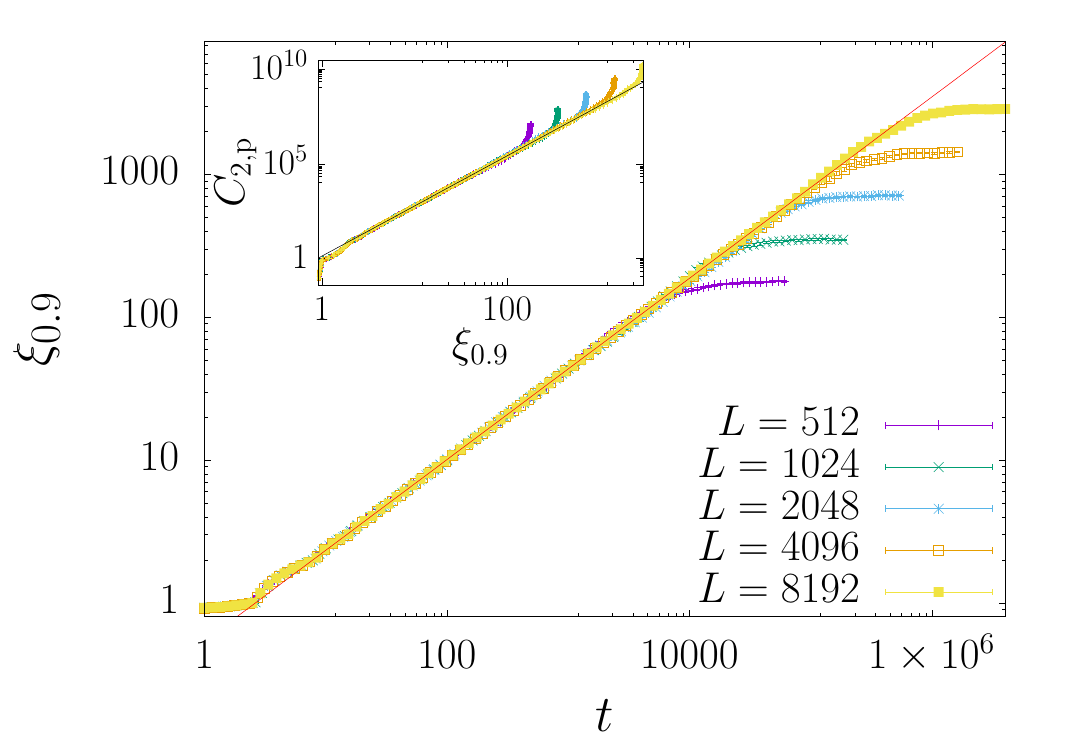}
    \includegraphics[width=0.33\textwidth]{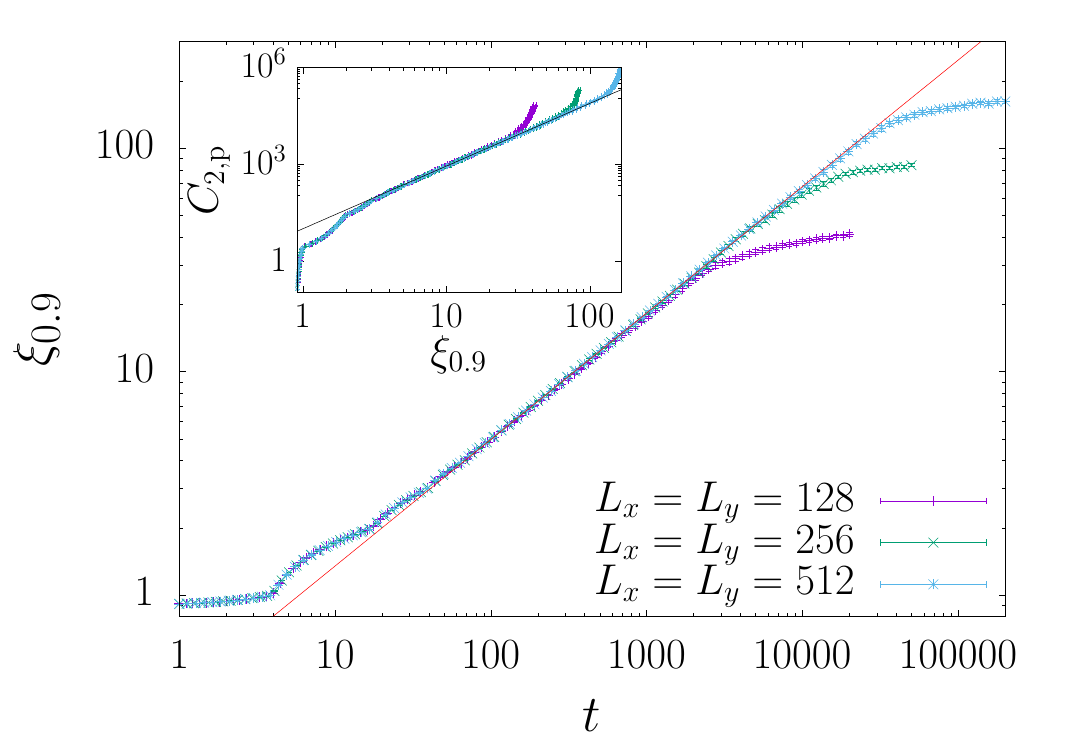}
    \includegraphics[width=0.33\textwidth]{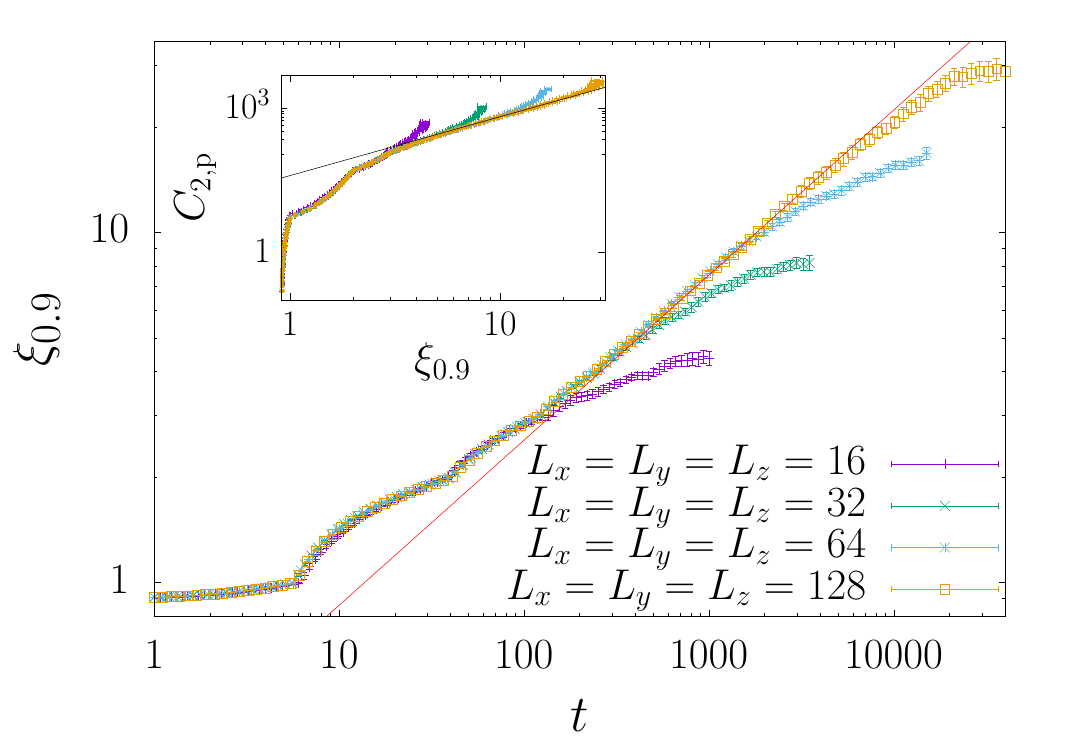}
    \caption{Correlation length $\xi_{0.9}(t)$ against $t$ for $d=1,2$ and 3 from left to right. Red solid lines correspond to the fit $t^{1/z}$ in each case. Insets: Correlation function at the plateau, $C_{2,p}(t)$ vs $\xi_{0.9}(t)$, for $d=1$, 2 and 3, left to right. Black solid lines correspond to the fit $C_{2,p} \sim \xi_{0.9}^{2\alpha}$.} 
    \label{fig:long}
\end{figure*}

\section{Results}
\label{sect:results}
In this section we report the results of our numerical simulations of the CP. In particular we will show the results regarding the evolution with time of the particle density, the interface height, the front roughness, the correlation length, and the behavior of $C_2$. Finally, we report the results concerning the universality of front fluctuations and the scaling properties of $C_1$. In each case, we provide results for the three values of $d$ considered herein. The reader can find full details of the numerical simulations in Appendix \ref{app:details}. Results for additional observables related with intrinsic anomalous scaling (some of which were addressed in Ref.\ \cite{Dickman2000}) are reported in appendices \ref{app:SF} and \ref{app:gradient}.

In order to compute the statistical errors, we have used the jackknife procedure throughout \cite{Young2015,Efron1982}; see in particular Appendix C in Ref.~\cite{Barreales2020} for additional details in a similar context, namely kinetic roughening of a discrete model. We have used the convention that the numbers in round brackets give the estimated uncertainty in the last digit(s). These error bars are represented in the graphics, although in many cases they are difficult to see.

\subsection{Density and roughness: exponents $\theta$ and $\beta$}
\label{sec:thetabeta}

We have computed the particle density $\rho(t)$ as a function of time for one, two, and three dimensional systems of different sizes (see the insets in Fig.\ \ref{fig:w2}). Fitting the data to the scaling law $\rho(t)\sim t^{-\theta}$ \cite{Dickman2000}, we can measure the exponent $\theta$; these results are shown in Table \ref{tab:exponents}.

Our results for $\theta$ and $\beta$ are consistent with the values and with the $\beta=1-\theta$ scaling relation reported in Ref.\ \cite{Dickman2000}; the agreement with the latter improves for increasing $L$. Indeed, as argued therein, 
given that in the CP $\rho(t) \sim t^{-\theta}$, the mean height of the interface $\overline{h}(t)$ obeys
\begin{equation}
\overline{h}=\int dt\,\rho \sim \int dt\, t^{-\theta} \sim t^{1-\theta}\,.
\end{equation}
The front roughness is defined as the standard deviation of the local height values, so that $w^2\sim \overline{h}^2$; comparison with the $w^2\sim t^{2\beta}$ kinetic roughening behavior thus implies $\beta=1-\theta$, {consistent with our numerical results. 

\begin{table*}[tb]
\caption{Exponents $\theta$, $\beta$, $z$, $\alpha$, and $\alpha_{\mathrm{loc}}$ for $d =$ 1, 2 and 3, as obtained in our simulations. The size of the system is $L^d$ in each case.}
\label{tab:exponents}
\begin{ruledtabular}
\begin{tabular}{lrlllll}
$d$ & $L$ & $\theta$ & $\beta$ & $z$ & $\alpha$ & $\alpha_{\mathrm{loc}}$\\
\colrule
1 & 512  & 0.1595(6)  & 0.8195(18) &  1.589(13)& 1.341(8) & 0.624(8)\\
  & 1024 & 0.1610(3)  & 0.825(2) & 1.587(8) & 1.336(6) & 0.631(6)\\
  & 2048 & 0.1608(2)  & 0.8317(17) & 1.608(10) &  1.343(7) & 0.636(8)\\
  & 4096 & 0.1607(3) & 0.8354(7) & 1.577(4) & 1.328(3) & 0.644(3)\\
  & 8192 & 0.1610(2)  & 0.8373(5) & 1.573(3) &  1.324(2) & 0.644(2)\\
\colrule
2 & 128 & 0.4489(4) & 0.5440(13) &  1.85(2) &  1.026(9) & 0.430(9)\\
  & 256 &  0.4515(7) & 0.5452(14) &  1.793(9) & 0.988(6) & 0.440(6)\\
  & 512 &  0.4518(4) & 0.5461(6) & 1.765(8) & 0.970(7) & 0.453(7)\\
\colrule
3 & 64 & 0.714(4) & 0.292(3) & 2.20(5) & 0.683(14) & 0.188(15)\\
  & 128 & 0.726(4) & 0.288(3) & 2.12(14) & 0.62(4) & 0.17(4)\\
\end{tabular}
\end{ruledtabular}
\end{table*}

\subsection{Front correlation length: exponents $\alpha$ and $z$}

To obtain the values of additional exponents, we next compute the correlation length $\xi_{0.9}(t)$ as the position $r$ for which the correlation function $C_2(r,t)$ reaches $90\%$ of its value at the plateau, as explained in Sec.\ \ref{sect:observables}. Once the correlation length is calculated, it can be plotted versus time and fitted to Eq.~(\ref{eq:corr_length}), to obtain the dynamic exponent $z$. Similarly, the height-difference correlation function at the plateau, $C_{2,p}(t)$, is plotted against the correlation length in order to measure the exponent $\alpha$ according to Eq.~(\ref{eq:c2p}).
Figure \ref{fig:long} shows the plots just mentioned, for $d=1, 2$, and $3$. The values of $z$ and $\alpha$ resulting from our fits for different dimensions and various system sizes are collected in Table \ref{tab:exponents}.

As a consistency check, recall from Sec.\ \ref{sect:observables} that $\alpha$, $\beta$, and $z$ satisfy an scaling relation which allows us to calculate the growth exponent as $\beta=\alpha/z$, directly computed from the values of $\alpha$ and $z$. This value of $\beta$ can be compared with that measured from the front roughness (denoted in this paragraph as $\beta_w$). For $d=1$ and $L=8192$, the value of $\beta=\alpha/z=0.842(2)$ is compatible with $\beta_w=0.8273(5)$ by two standard deviations. Moreover, for $d=2$ and $L=512$, $\beta=\alpha/z=0.550(5)$ is compatible with $\beta_w=0.5461(6)$ in the uncertainty interval. And finally, for $d=3$ and $L=128$, $\beta=\alpha/z=0.29(3)$ is in good agreement with $\beta_w=0.288(3)$. Overall, the exponent values we obtain are consistent with those reported in Ref.\ \cite{Dickman2000} for all $d$, the largest differences occurring for $d=3$, where our value for $\alpha$ ($z$) is 2.9 (1.5) standard deviations away from that reported therein. 

\subsection{Height-difference correlation function}
\label{sect:c2}

\begin{figure*}[htb]
    \centering
    \includegraphics[width=0.33\textwidth]{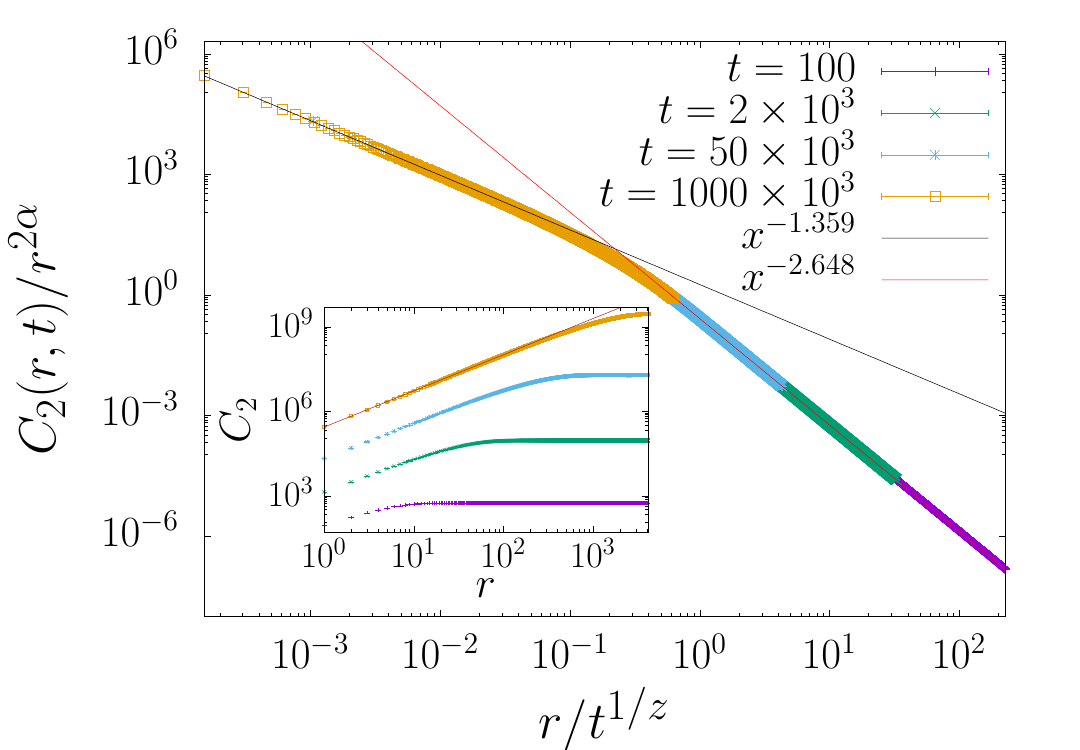}
    \includegraphics[width=0.33\textwidth]{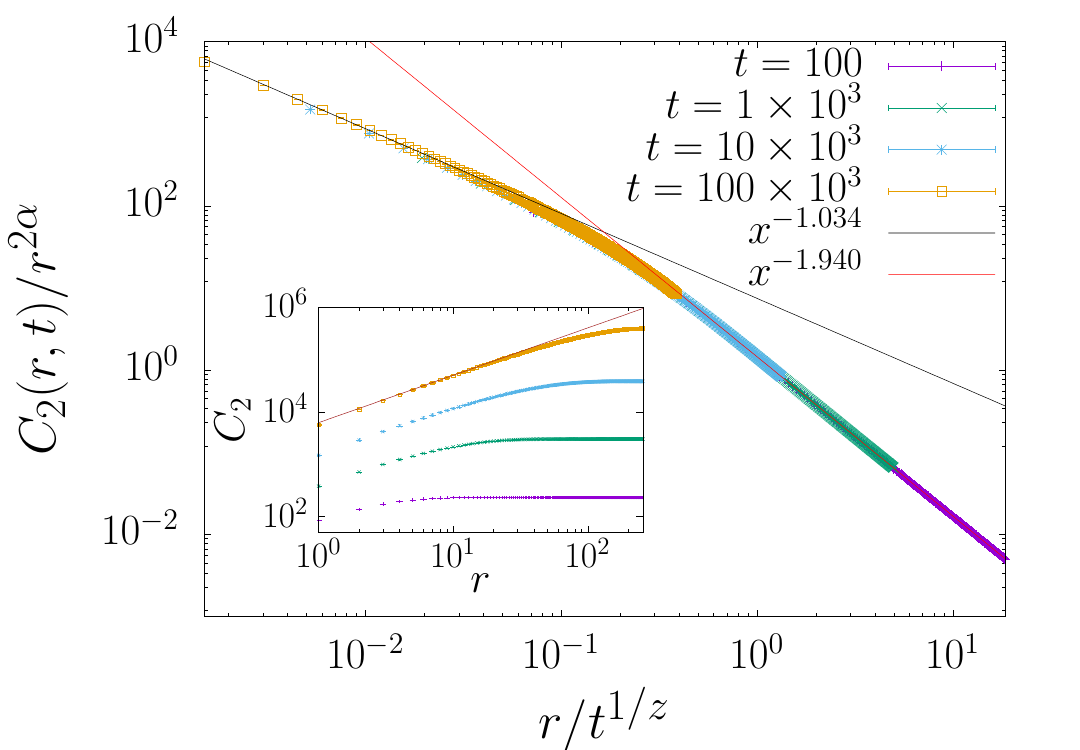}
    \includegraphics[width=0.33\textwidth]{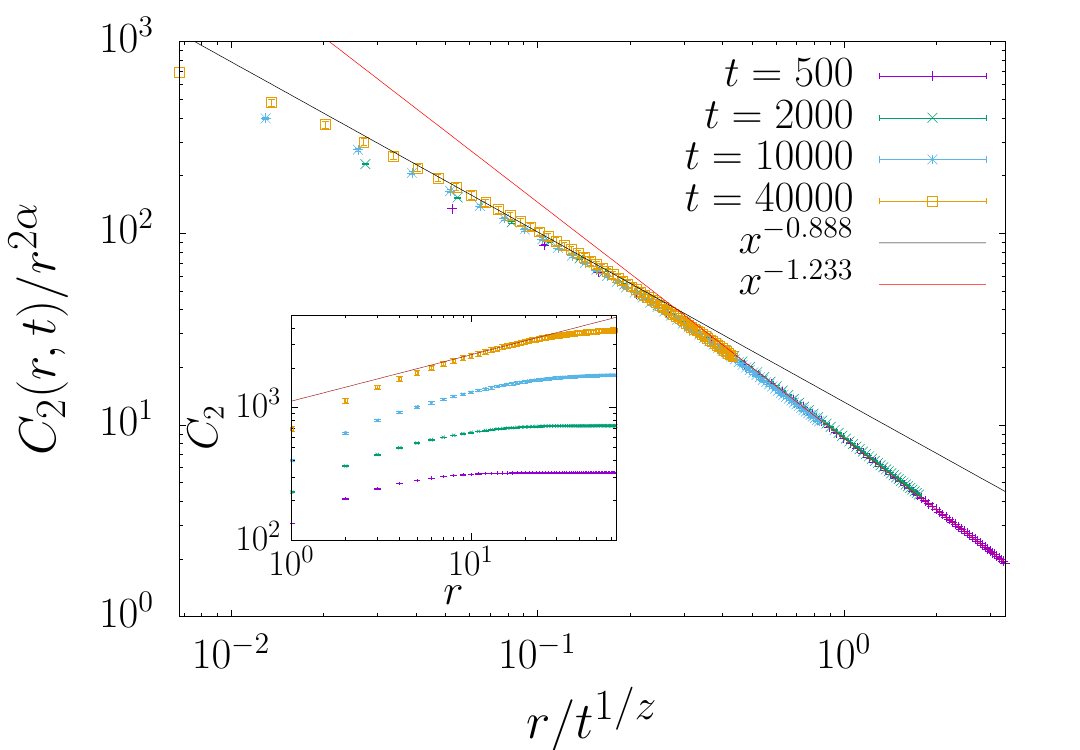}
    \caption{Data collapse of $C_2(r,t)$ for $d=1$, 2 and 3 and sizes $L=8192$, $512$, and $128$, respectively, for different values of time, as indicated in each legend. The exponents $\alpha$ and $z$ are those shown in Table \ref{tab:exponents}. Solid lines are proportional to $x^{-2\alpha'}$ and $x^{-2\alpha}$. Insets: Uncollapsed $C_2(r,t)$ data as a function of $r$ for the same conditions as in the corresponding main panel. Solid lines are proportional to $r^{2\alpha_{\rm loc}}$, where the exponents $\alpha_{\rm loc}$ are those of Table \ref{tab:exponents}.}
    \label{fig:c2}
\end{figure*}

Information on the local scaling behavior of the front is provided by the full height-difference correlation function $C_2(r,t)$, which has been likewise computed in 1, 2, and 3 dimensions. The insets in Fig.\ \ref{fig:c2} show $C_2(r,t)$ as a function of $r$ for several values of $t$ and for the various $d$, choosing the largest system size $L$ for each dimension. We observe that, irrespective of $d$, the $C_2(r,t)$ curves obtained for different times shift systematically upwards with increasing time and do not overlap for any value of $r$. This fact implies the occurrence of anomalous scaling, which, in principle, can be originated by different causes as noted in Sec.\ \ref{sect:introduction}. In this case, we can measure an additional roughness exponent, $\alpha_{\rm loc}$. We have represented $C_2(r,t)/r^{2\alpha}$ vs $r/t^{1/z}$ using our estimates of $\alpha$ and $z$, see Fig.\ \ref{fig:c2}. 
According to Eq.~(\ref{eq:c2an}), $\alpha'=\alpha-\alpha_{\rm loc}$ may be estimated for small arguments of the scaling function, from which we obtain the value of $\alpha_{\rm loc}$ shown in Table \ref{tab:exponents}. The values of $\alpha_{\rm loc}$ differ slightly by varying $L$, but we interpret the differences as due to our finite systems sizes. The fact that $\alpha \neq \alpha_{\rm loc}$ while $\alpha_{\rm loc}<1$ qualifies the present type of behavior as intrinsic anomalous scaling \cite{Lopez1997,Ramasco2000,Cuerno2004}. 

Note that $\alpha$ takes relatively large values for all $d$. In these cases, in particular in the presence of anomalous scaling \cite{Lopez1997,Lopez1997b}, two-point correlations are frequently studied in Fourier space~\cite{Siegert1996}. In Appendix \ref{app:SF} we present an analysis of our data based in the study of the front structure factor \cite{Barabasi1995,Krug1997}, which reaches the same conclusions on the anomalous scaling of CP interfaces.
Likewise, Appendix \ref{app:gradient} contains another consistency check on the intrinsic anomalous scaling that is found in our simulations. Namely, we verify the scaling law for the time evolution of the average surface slope which is expected in this context \cite{Lopez1999} and was also verified in the simulations of Ref.\ \cite{Dickman2000}.

\subsection{PDF of front fluctuations}

Beyond scaling exponent values, we consider the statistics of front fluctuations. Specifically, we represent the histogram of the front fluctuations rescaled as in Eq.~(\ref{eq:fluc}). To evaluate this histogram, for each dimension,  we evaluate $\chi(t)$ at values of time within an interval in which the front roughness scales as $w(t) \sim t^\beta$,  with the growth exponent value which was discussed in Sec.~\ref{sec:thetabeta}.
Figure \ref{fig:PDF} plots the front fluctuation histogram for one, two, and three dimensions. We note that $P(\chi)$ is independent of $L$ within the statistical precision in all cases. For comparison, the figure shows the exact PDF for the Gaussian case, that is found for linear models of kinetically rough interfaces such as the EW equation with time-dependent noise \cite{Barabasi1995,Krug1997,Carrasco2019}, and the PDF for the KPZ universality class, which is dimension-dependent. Notice we have normalized our data to zero-mean and unit-variance.

In $d=1$, the fluctuation PDF for rough interfaces in the $d=1$ KPZ universality class using periodic boundary conditions is provided by the Tracy-Widom (TW) distribution for the largest eigenvalue of Hermitian random matrices in the Gaussian orthogonal ensemble (GOE-TW) \cite{kriecherbauer2010,Takeuchi2018}. 
For $d>2$, distributions other than TW play analogous roles to the latter for the KPZ universality class \cite{Halpin-Healy2012,Halpin-Healy2013}. For $d=2$ we show in Fig.\ \ref{fig:PDF} the distribution obtained for an Euler integration of the KPZ equation and reported in Ref.\ \cite{Halpin-Healy2012}. Finally, for $d=3$ we show the fluctuation PDF for the 3D radial KPZ class from the DPRM/SHE data collapse performed in Ref.\ \cite{Halpin-Healy2013}. 

While the numerical PDFs we obtain for the CP are certainly non-symmetric (hence, with non-zero skewness as for the KPZ case) for all the simulated dimensions, they all differ appreciably from these distributions. Moreover, they exhibit a strong dependence with the dimensionality of the system. In Fig.~\ref{fig:flu_D} we have represented the distributions for each of the simulated dimensions together, with the data being available at the Zenodo open access repository \cite{datasetchi}. The corresponding skewness $s$ and excess kurtosis $k$ values for the largest systems are collected in Table \ref{tab:skewness}. Notably, the signs of $s$ and $k$ for $d=1$ differ from their values for $d=2,3$, again at variance with the KPZ universality class. However, the magnitudes of $s$ and $k$ do increase with $d$, as is the case also for KPZ systems \cite{Halpin-Healy2012,Halpin-Healy2013,Alves2014}.

Let us now focus on the tails of the PDFs. These tails are asymmetric and non-Gaussian, and are seen to follow exponential functions for many random systems in their disorder-dominated phases  \cite{Monthus2008} as
\begin{equation}
    P(x) \approx \left\{ \begin{array}{lc}
             e^{-c|x|^{\eta_-}},  & x \rightarrow -\infty  \\
             \\ e^{-dx^{\eta_+}}, & x \rightarrow +\infty, \\
             \end{array}
   \right.
   \label{eq:tails}
\end{equation}
where $c$ and $d$ are constants and $\eta_-$ and $\eta_+$ are characteristic so-called tail exponents \cite{Monthus2008,Halpin-Healy2015}. We have fitted the distributions computed herein to functions \eqref{eq:tails} (dashed lines in Fig.~\ref{fig:PDF}) to estimate the $\eta_-$ and $\eta_+$ exponents; the results are collected in Table \ref{tab:skewness}. The tail exponents for the Tracy-Widom distribution have been studied in detail \cite{majumdar2014,Halpin-Healy2015}. For the $d=1$ KPZ universality class, the exponent of the Airy tail (i.e., right tail in the GOE-TW representation of Fig.\ \ref{fig:PDF}) is $\eta_+=3/2$, whereas for the left tail one has $\eta_-=3$ \cite{kim1991, majumdar2014}. In addition, for KPZ the left and right tail exponents are related to each other as $\eta_-=(d+1)\eta_+$, and also to the growth exponent through $\eta_+=1/(1-\beta)$ \cite{majumdar2014,Halpin-Healy2015}. Note, however, that although the fits are reasonable (specially for $\eta_+$) and such that $\eta_->\eta_+>1$ for all $d$ as implied by the KPZ formulae, our data do not suggest any simple connection between the right and left tail exponents, or between them and $\beta$. Departure from the $\eta_+=1/(1-\beta)$ relation can be found elsewhere, e.g.\ for synchronized oscillator lattices \cite{Gutierrez2023}, for which the PDF is GOE-TW in spite of the fact that the kinetic roughening exponents ($\beta$, in particular) do not take their 1D-KPZ values.

\begin{figure*}[htb]
    \centering
    \includegraphics[width=0.33\textwidth]{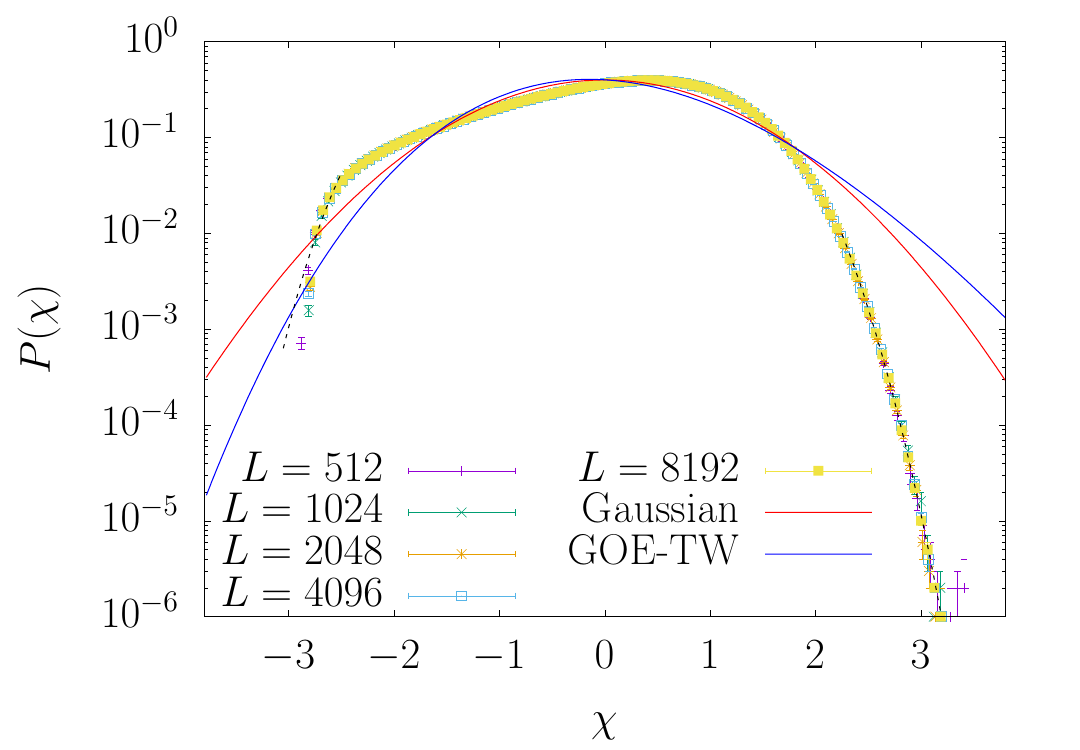}
    \includegraphics[width=0.33\textwidth]{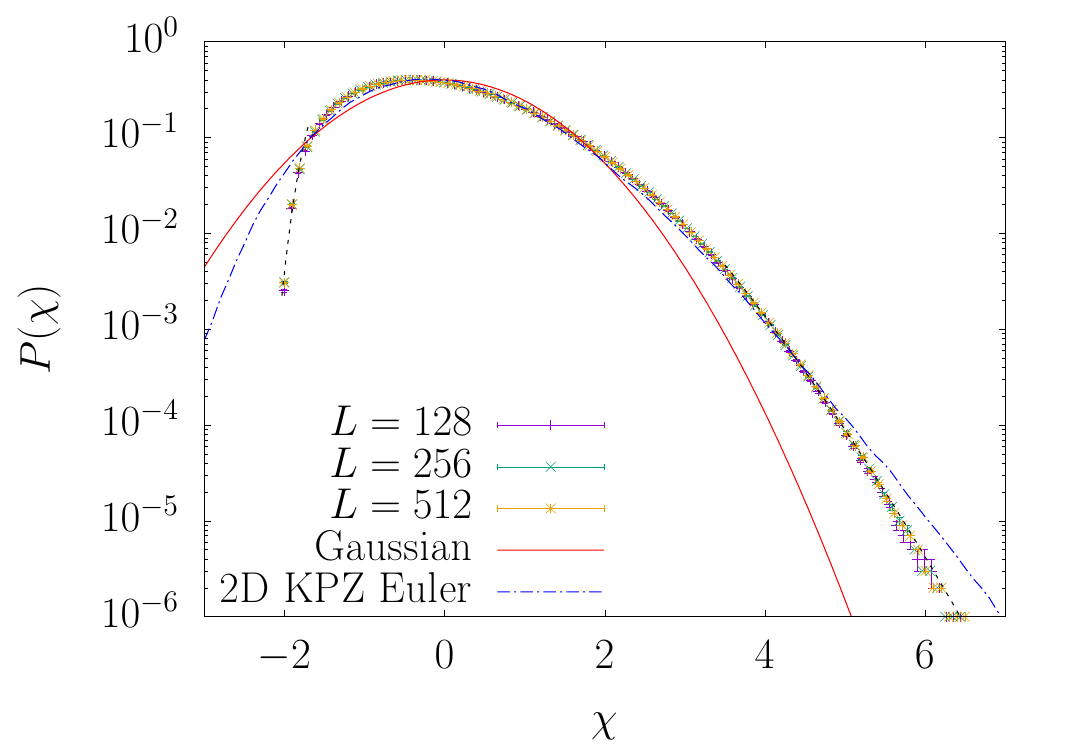}
    \includegraphics[width=0.33\textwidth]{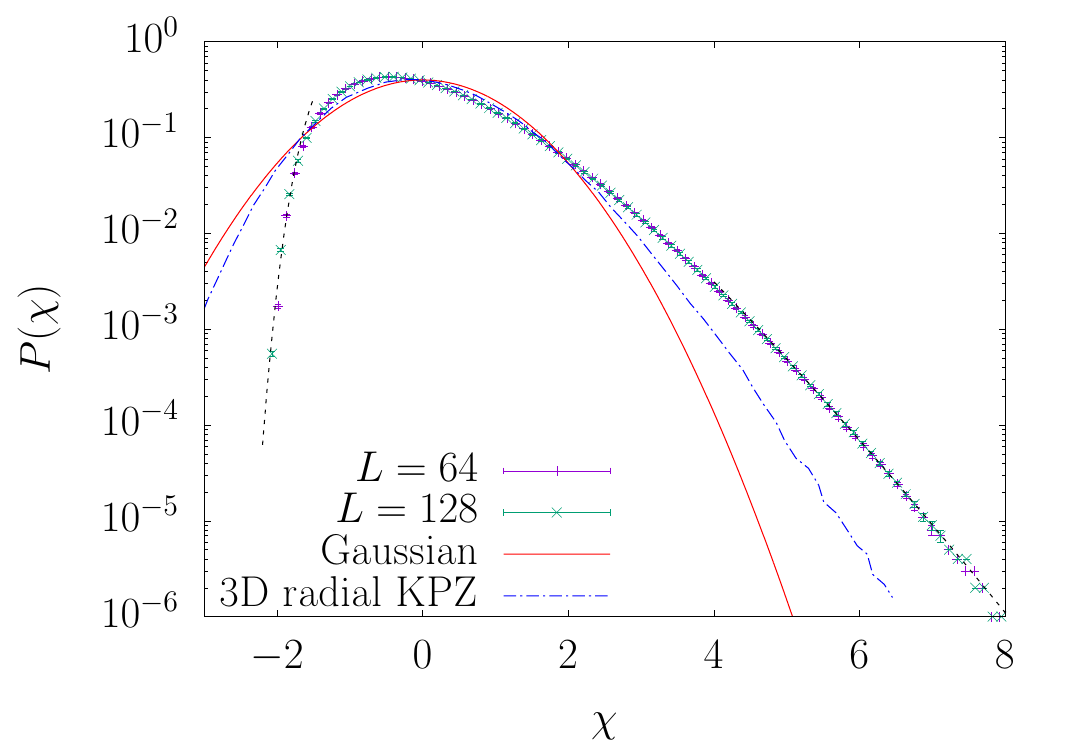}
    \caption{PDF of front height fluctuations for $d=1, 2,$ and 3 (left to right panels), and sizes indicated in each legend. For comparison, the solid lines correspond to the exact Gaussian distribution, the GOE-TW distribution, the $d=2$ KPZ fluctuation distribution (Euler integration \cite{Halpin-Healy2012}) and the $d=3$ KPZ fluctuation distribution (radial KPZ class \cite{Halpin-Healy2013}), as described in the legends. The dashed black lines correspond to the tail functions (left) $e^{-c|x|^{\eta_-}}$ and (right) $e^{-x^{\eta_+}}$, with $\eta_-$ and $\eta_+$ exponent values as reported in Table \ref{tab:skewness}.}
    \label{fig:PDF}
\end{figure*}

\begin{figure}[htb]
    \centering
    \includegraphics[width=0.45\textwidth]{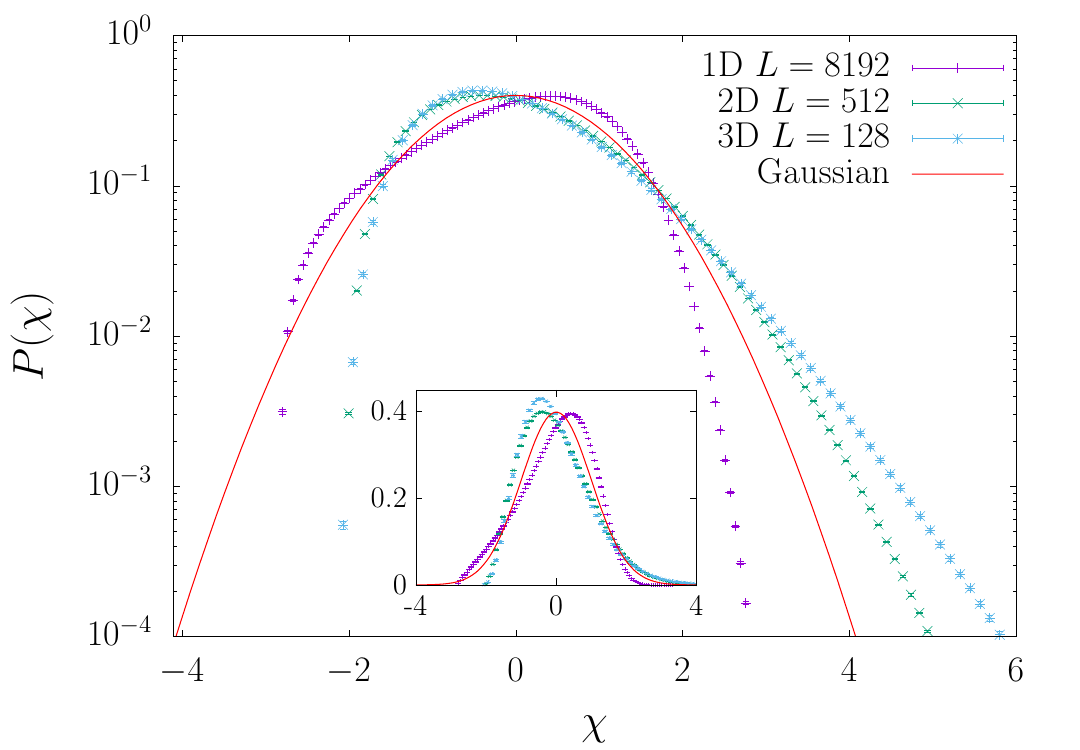}
    \caption{Histograms of front fluctuations in different dimensions. We represent probabilities $P>10^{-4}$ for greater visibility and a better comparison between the different dimensions. The solid lines show the exact Gaussian distribution. Inset: Same data in linear scale. The numerical data for this figure are openly available at Ref.\ \cite{datasetchi}.}
    \label{fig:flu_D}
\end{figure}

\begin{table}[htb]
\caption{Skewness $s$, excess kurtosis $k$, and tail exponents $\eta_-$, $\eta_+$ for the front fluctuation distributions in different dimensions, as obtained in our simulations for the largest value of $L$ in each case.}
\label{tab:skewness}
\begin{ruledtabular}
\begin{tabular}{cccccc}
$d$ & $L$ & {$s$} & $k$ & $\eta_-$ & $\eta_+$ \\ 
\colrule
1  & 8192 & $-0.3677(15)$ & $-0.421(3)$ & $4.05(3)$ & $3.12(2)$ \\ 
\colrule
2  & 512 & $0.6246(18)$ & $0.233(3)$ & $6.261(15)$ & $1.559(6)$ \\ 
\colrule
3 & 128 & $0.848(5)$ & $0.987(17)$ & $5.12(2)$ & $1.251(13)$ \\ 
\end{tabular}
\end{ruledtabular}
\end{table}

\subsection{Front covariance}

\begin{figure*}[htb]
    \centering
    \includegraphics[width=0.33\textwidth]{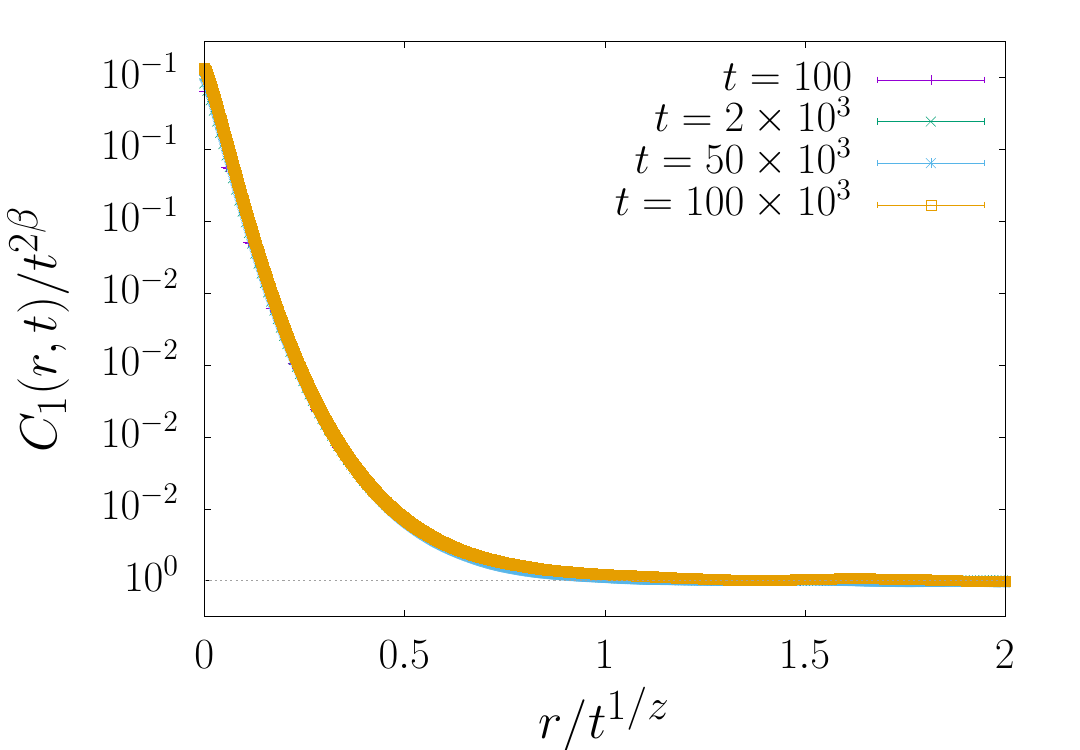} 
    \includegraphics[width=0.33\textwidth]{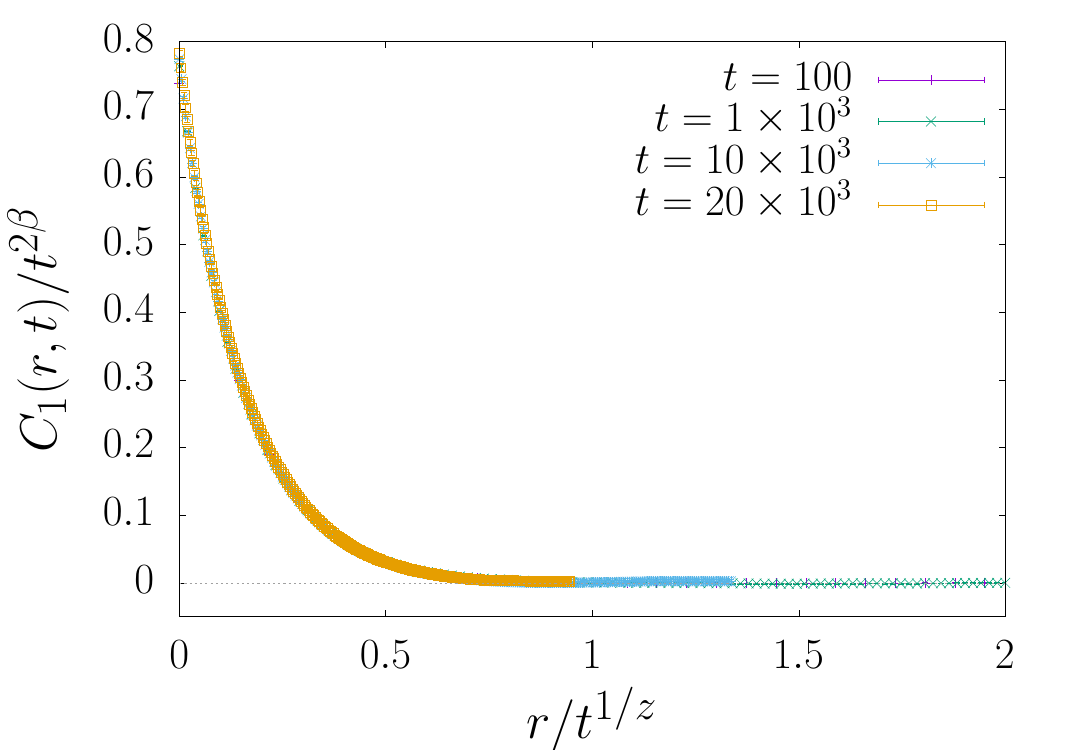} 
    \includegraphics[width=0.33\textwidth]{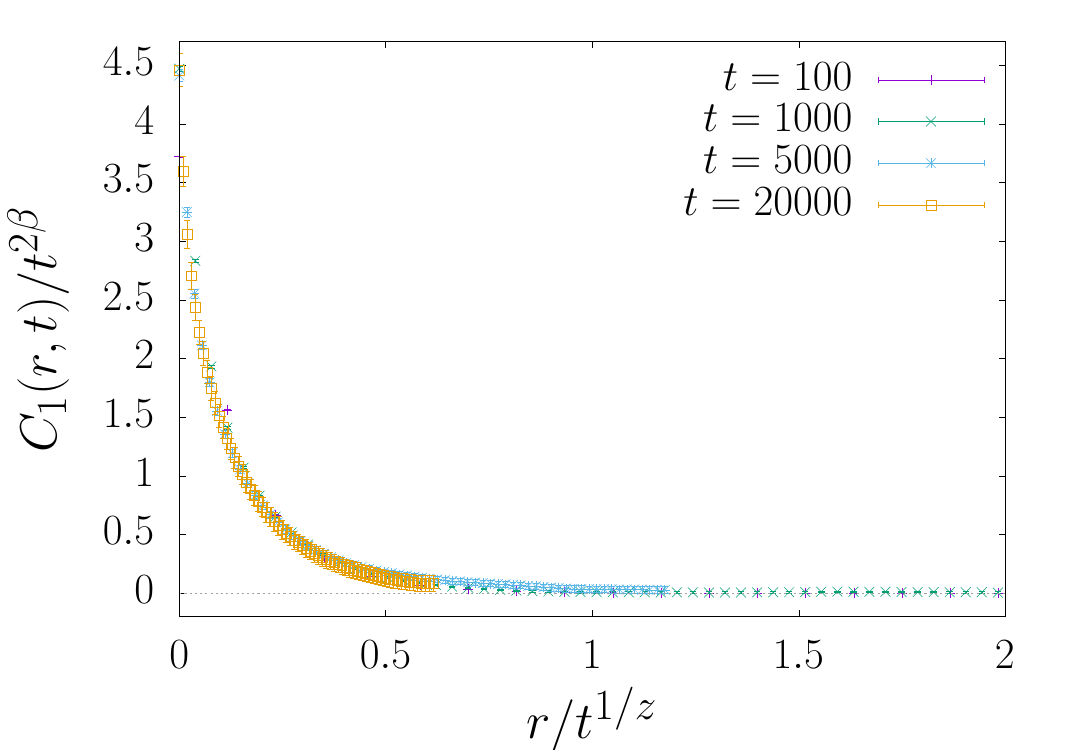}
    \caption{Scaled front covariance $C_1(r,t)$ for $d=$ 1, 2 and 3 and sizes $L=8192$, $512$ and $128$, respectively, left to right, at times as given in the legends. The exponents $\beta$ and $z$ are those shown in Table \ref{tab:exponents}. In each case, times larger than those shown do not scale properly.
    The numerical data for this figure are openly available at Ref.\ \cite{datasetC1}.}
    \label{fig:c1}
\end{figure*}

As noted in Sec.\ \ref{sect:introduction}, for kinetic roughening systems the front covariance correlation function $C_1(\boldsymbol{r},t)$, 
Eq.~(\ref{eq:correlation_1}), is expected to behave as 
\begin{equation}
    C_1(\boldsymbol{r},t)=t^{2\beta} F\big(r/t^{1/z}\big)\,,
    \label{eq:C1F}
\end{equation}
where $F(u)$ is an universal function which becomes an additional (albeit nonexclusive) trait of the universality class. For instance, in the one-dimensional (1D) KPZ case, $F$ is the so-called Airy$_n$ function, with $n=1$ or 2 depending on the boundary conditions \cite{Takeuchi2018}, a feature which happens to be shared by the 1D EW universality class \cite{Carrasco2019}. Likewise for the covariance of the 1D EW and KPZ equations with columnar noise, recently found to be provided in both cases by that of the Larkin model of elastic interfaces in disordered media \cite{Gutierrez2023}.  

Here we assess the universal scaling of the covariance, Eq.\ \eqref{eq:C1F}, for the CP in dimensions $d=1, 2$, and 3. We have represented this scaling Ansatz using our values for $\beta$ and $z$ in each case. Figure \ref{fig:c1} shows the results for the largest size for $d=1$ ($L=8192$), $d=2$ ($L=512$), and $d=3$ ($L=128$). As expected, we note that the rescaled curves for different times do overlap; an analogous behavior is obtained if we represent together curves for different system sizes. Note, moreover, that the exponents entering this collapse ($\beta$ and $z$) are global ones, even in a case like this in which scaling is intrinsically anomalous, i.e.\ non-FV. For each system size, there exists a maximum time above which curves for different times do not overlap; but this is a finite-size effect, since this maximum time increases systematically with $L$. 

The CP scaling function changes quantitatively with $d$, although its qualitative behavior does not. In particular, the $d=1$ case is worth considering in detail. As noted above, for systems in the 1D KPZ universality class with periodic boundary conditions, the height covariance $C_1(r,t)$ behaves as
\begin{equation}
    C_1(r,t)=a_1 \, t^{2\beta} \mathrm{Airy}_1\left(a_2 r/t^{1/z} \right) \,,
   \label{eq:airy1}
\end{equation}
where ${\rm Airy}_1(u)$ denotes the covariance of the ${\rm Airy}_1$ process \cite{Halpin-Healy2015, Takeuchi2018, bornemann2010}, and $a_1$ and $a_2$ are fitting constants. Our numerical data for $C_1(r,t)$ in $d=1$ seem to agree with Eq.\ \eqref{eq:airy1}, as shown in Fig.\ \ref{fig:1Dc1Airy}. While admittedly quantitative differences exist for small values of $r/t^{1/z}$ (see the inset of Fig.\ \ref{fig:1Dc1Airy}), the relative error between the theoretical and the numerical curves in this region is not larger than $3\%$.

In two and three dimensions, our numerical CP data for $C_1(\boldsymbol{r},t)$  do not scale with ${\rm Airy}_1(u)$ (not shown), and are not expected to as, e.g.\ for the KPZ class itself, Airy$_n$ behavior seems to be specific of the one dimensional case. Our data for 2D do not seem to agree with the 2D $KPZ$ universality class either, whose covariance is numerically well-characterized \cite{Halpin-Healy2014,Almeida2014}. The CP front covariances, for each dimension, are also available as open data from \cite{datasetC1}.

\begin{figure}[hbt]
    \centering
    \includegraphics[width=0.45\textwidth]{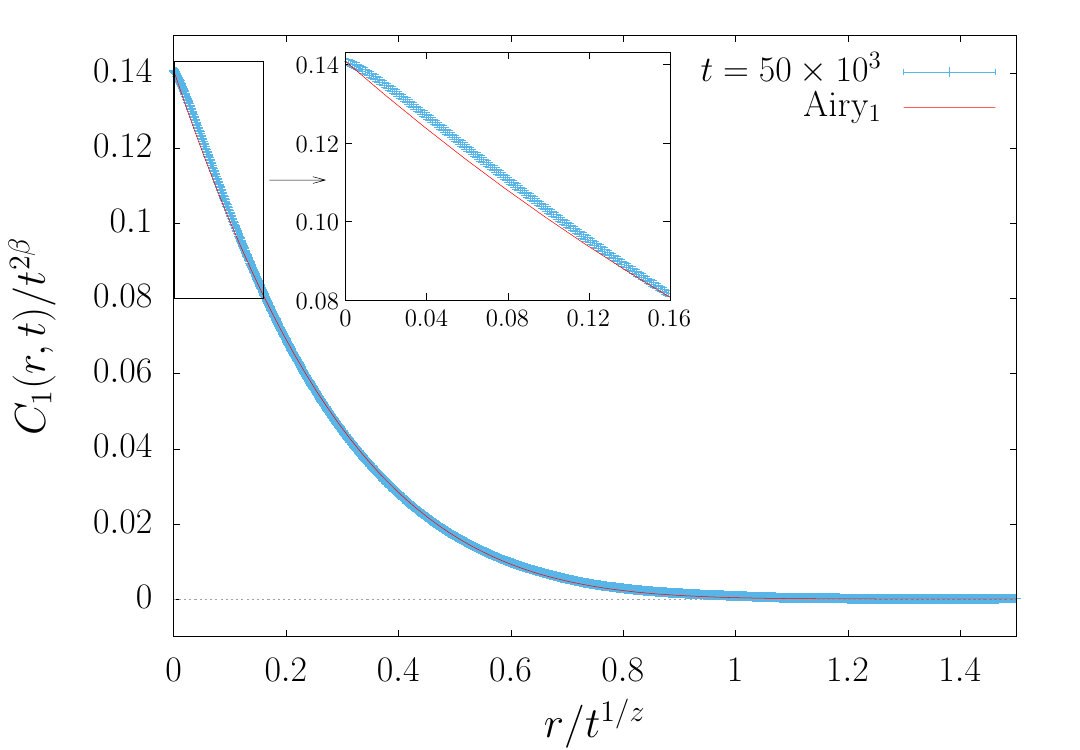}
    \caption{The scaled CP front covariance $C_1(r,t)$ for $d =$ 1 and $L=8192$ at $t=5\times 10^4$. The solid line shows a function proportional to the covariance of the Airy$_1$ process. Inset: Zoom of the boxed small-argument region in the main panel. The numerical data for this figure are openly available at Ref.\ \cite{datasetC1}.}
    \label{fig:1Dc1Airy}
\end{figure}

\section{Discussion}
\label{sect:disc}

We have obtained that, working at the absorbing state phase transition, the interface problem associated with the CP displays the full array of traits of a kinetic roughening universality class: well defined, $d$-dependent scaling exponents, fluctuation PDF, and covariance, in such a way that a dynamic scaling Ansatz, which happens to be intrinsically anomalous, is consistently satisfied.

Note at this that the continuum description of the CP as a particle model is provided by the so-called Reggeon field theory, which, in suitable units, corresponds to the following reaction-diffusion equation for a {\em local} density field $\rho(\boldsymbol{x},t)$ \cite{Henkel2008},
\begin{align}
    \partial_t \rho &=D\nabla^2 \rho + \rho - \rho^2 + \sqrt{C \rho} \; \eta(\boldsymbol{x},t) ,
    \label{eq:RFT}
\end{align}
where $D, C >0$ are constants and $\eta(\boldsymbol{x},t)$ is a zero-average Gaussian noise of unit variance. 
Note the absorbing nature of the $\rho=0$ state, which suppresses both dynamics and fluctuations. We find it interesting that the kinetic roughening universality class of the (interface representation of the) CP turns out to be so different as compared with that of systems described by the (deceivingly) similar stochastic Fisher-Kolmogorov-Petrovsky-Piscounov (sFKPP) equation, which reads
\begin{equation}
    \partial_t \rho=D\nabla^2 \rho + \rho - \rho^2 + \sqrt{\rho(1-\rho)/N} \; \eta(\bm{x},t) ,
    \label{eq:stoch_fisher}
\end{equation}
where $N>0$ is a constant and $\eta(\bm{r},t)$ is the same noise as in Eq.\ \eqref{eq:RFT}. Indeed, Eq.\ \eqref{eq:stoch_fisher} is known to provide a coarse-grained description for the $A+A \leftrightarrow A$ reaction-diffusion particle model \cite{Ben-Avraham1990,Pechenik1999,Doering2003}. But remarkably, in this case the kinetic roughening behavior of the corresponding moving front problem is in the standard KPZ universality class \cite{Nesic2014,Barreales2020}. Quite possibly the DP behavior at the absorbing phase transition in the CP \cite{Odor2004,Henkel2008} is at the core of this stark difference between Eqs.\ \eqref{eq:RFT} and \eqref{eq:stoch_fisher}, and likewise for their corresponding particle models. From this point of view, the existence of absorbing states in particle models may play a similar role in the (anomalous) kinetic roughening scaling Ansatz of interface systems they map into to that played by morphological instabilities and/or quenched disorder \cite{Lopez2005}.

Still, we find it interesting that, at least in the 1D case, the CP and the standard KPZ (and EW) universality classes seem to share (to a good precision) the same Airy covariance. Another intriguing similarity of the interface representation of the CP is with the Mullins-Herring equation with quenched disorder \cite{Song2006}. Indeed, not only has this model been seen to feature anomalous scaling \cite{Song2008}, but its scaling exponents (including those characterizing the pinning transition seen to occur) are numerically quite close to those of CP/DP, at least for $d=1$ and 2 \cite{Song2011}. Recall at this that DP is also known to control the interface scaling at depinning for other well known interfacial systems, such as the KPZ equation with quenched disorder \cite{Barabasi1995,Wiese2022,Barreales2022}, for which the universality class is in turn different from the one we are presently studying in this paper.

With respect to the behavior of anomalous kinetic roughening with dimension, note that recent observations of its occurrence in the synchronization of oscillator lattices \cite{Gutierrez2023} enhance its potential interest, due to the high connectivity (thus, a large effective dimensionality) of synchronizing agents in many applications \cite{Pikovsky2001,Arenas2008}. For the CP we obtain a consistent intrinsic anomalous scaling Ansatz for $d<d_c$, which is further endowed with an universal PDF and covariance for its front fluctuations. For increasing $d$, both $\alpha$ and $\alpha_{\rm loc}$ decrease (note that the saturation roughness $w_{\rm sat}$ does not scale with $L$ for $d\geq d_c$), as expected \cite{Barabasi1995}. However, the relative difference $(\alpha-\alpha_{\rm loc})/\alpha$ does not become particularly reduced with increasing $d$, so that the intrinsic anomaly in the scaling persists somehow all the way up to the upper critical dimension. On the other hand, with increasing $d$ our numerical values for the dynamic exponent do increase towards the diffusive $z=2$ value expected \cite{Dickman2000} at and above $d_c$, with some overshoot in the 3D case. This fact could be a size effect. Indeed, Table \ref{tab:exponents} exhibits a dependence of all exponents on the system size, which is particularly noticeable for $d =$3. In this sense, we cannot guarantee that our values for $L = 128$ may be considered as asymptotic, especially for exponents $z$ and $\alpha$. Note that exponents $\theta$ and $\beta$ for $d = 3$ are in good agreement with theoretical predictions, though.

An interesting question is how the intrinsic anomalous scaling would emerge in a putative continuum description of the interface problem associated with the CP, and actually how such a model would look like. Recall that intrinsic anomalous scaling has been conjectured not to be asymptotic for systems with local interactions in which neither morphological instabilities nor quenched disorder occur \cite{Lopez2005}; however, ways out of this prediction exist, as for the tensionless KPZ equation \cite{Rodriguez-Fernandez2022}, possibly through non-perturbative behavior. Recall also that various discrete models are known which feature intrinsic anomalous scaling, while their continuum limits do not, see Ref.\ \cite{Marcos2022} for a recent discussion.

With respect to the universal fluctuation PDF and covariance computed herein for the first time, we would like to stress the strong non-Gaussian, dimension-dependent features of the former.
In the context of kinetic roughening systems, to date non-Gaussian behavior has been reported either for the KPZ universality class, \cite{kriecherbauer2010,Halpin-Healy2015,Takeuchi2018} 
or for other classes somehow related with it (some of which feature anomalous scaling), like those of the conserved KPZ equation \cite{Carrasco2016}, the tensionless KPZ equation \cite{Rodriguez-Fernandez2022}, precursor spreading \cite{Marcos2022}, or systems related with the KPZ equation with columnar disorder \cite{Gutierrez2023}. From this point of view, the CP turns out to be innovative by providing alternative avenues for non-Gaussian interfacial behavior. Regarding the covariance, the 1D CP moreover provides another non-trivial example of a system with Airy behavior in spite of not having KPZ exponents, adding to previously reported cases \cite{Carrasco2019,Marcos2022}. 
Since these universal functions depend on the initial conditions and/or the boundary conditions in the KPZ case, we believe that it would be very interesting to study the CP interface model with different (non-periodic) boundary conditions in order to determine if such modifications have an impact on the CP/DP universality class.
The data of the PDF of the rescaled height fluctuations and the $C_1$ correlation function, for the different dimensions, have been published in open access in Zenodo \cite{datasetchi, datasetC1}.  Hopefully all these behaviors may become integrated in future into a comprehensive theory of critical dynamics far from equilibrium. 

As mentioned in Sec.~\ref{sect:introduction}, relatively recent works demonstrate experimental evidence for the DP universality class in different dimensions. Lemoult {\it et al.} \cite{Lemoult2016} study the onset of turbulence in a Couette flow using DP to explain the transition scenario, which in this case is governed by the Reynolds number. They measure critical exponents in a one-dimensional system ($d=1$), which are collected in Table \ref{tab:exp} and show a good agreement with those of DP, recall Tables \ref{tab:DP} and \ref{tab:exponents}. Likewise, in Refs. \cite{Takeuchi2007,Takeuchi2009} the critical behavior of the transition between two topologically different turbulent states in nematic liquid crystals has been studied, yielding a complete set of static critical exponents in full agreement with those of the DP class in $d=2$, see Tables ~\ref{tab:DP} and \ref{tab:exp}. Compared to our results in Table \ref{tab:exponents}, with which they are fully compatible, we note that the uncertainty in the experimental measurements is significantly higher. 
All this suggests the interest of further studies, both theoretical and experimental, to fully characterize the universality of the contact process in view of the results of our present work.

\begin{table}[b]
\caption{Critical exponents measured in experimental realizations of DP for $d=1$ and 2. The exponents with an asterisk have been derived from other exponents of the corresponding reference.}
\label{tab:exp}
\begin{ruledtabular}
\begin{tabular}{lccc}
 & $d$ & $\theta$ & $z$ \\
\colrule
Turbulence in Couette flow \cite{Lemoult2016} & 1 & 0.16(2)* & 1.8(4)* \\ 
Turbulent liquid crystals \cite{Takeuchi2009} & 2 & 0.48(5) & 1.7(3)* \\
\end{tabular}
\end{ruledtabular}
\end{table}

\section{Summary and Conclusions}
\label{sect:summ}

We have numerically studied the interface representation of the CP particle model in one, two, and three dimensions computing the critical exponents ($\alpha$, $\alpha_{\rm loc}$, $\beta$, and $z$) and the statistical properties related with the universal fluctuations of the front.

As stated in the Introduction, the critical exponents do not fully characterize the universality class of the model; one must also study the local statistical properties of the front, namely the PDF of the rescaled height fluctuations ($\chi$) and the scaling of the $C_1$ correlation function, which provides the front covariance.

We have presented a detailed analysis of the statistics of the local front fluctuations and that of $C_1$ to provide the additional, missing information needed to fully characterize the DP (or CP) universality class in all the physical dimensions below the upper critical one.

We have found that the PDF of the local fluctuations of the front does not follow any previously reported behavior (e.g., GOE-TW or Gaussian) instead exhibiting a strong dependence with the dimensionality of the system. There is an unexpected change of shape between the distribution for $d=1$ and those for $d=2,3$, with a change of sign of both skewness and kurtosis.

The front covariance exhibits a similar qualitative behavior when the dimension of the system changes, but its quantitative behavior is certainly different, again strongly dependent on the dimensionality of the system. In particular we have found that the covariance of the 1D fronts mimics that of the $\mathrm{Airy}_1$ process with high accuracy (of about $3\%$), but we have been unable to associate known analytical functional forms in any other of the simulated dimensions.

In addition, we have recomputed all the critical exponents of the CP model in these dimensions; in particular, we have estimated the dynamic critical exponent by computing directly the correlation length via the analysis of the heigth-difference correlation function $C_2$ in real space. 

Moreover we have thoroughly studied the intrinsic anomalous scaling displayed by this model, both in real and in Fourier space, allowing us to compute the local roughness exponent.

We have explicitly shown, in different plots, the behavior of the most important observables (and in some cases its associated scaling properties) for the three simulated dimensions. Overall, we have found a good agreement with the exponents previously reported in the literature \cite{Dickman2000}.

The associated statistical uncertainties have been thoroughly computed (for all the observables and PDF reported in the text) using the jackknife method in order to cope with the extremely strong correlations of the data. Without this methodology, the standard fit procedures (based in a diagonal $\chi^2$ analysis, i.e., neglecting completely the  correlation among the data) underestimate the statistical errors by more than a factor 10.

Finally, we consider that with the numerical characterization of the statistical  fluctuation properties of the front in one, two, and three dimensions, we have provided important pieces of information which were lacking, being needed to fully characterize the kinetic roughening behavior of one of the most important non-equilibrium universality classes, that of Directed Percolation.

\begin{acknowledgments}
This work was partially supported by Ministerio de Ciencia, Innovaci\'on y Universidades (Spain), Agencia Estatal de Investigaci\'on (AEI, Spain, 10.13039/501100011033), and European Regional Development Fund (ERDF, A way of making Europe) through Grants Nos.\ PID2020-112936GB-I00, PGC2018-094763-B-I00, and PID2021-123969NB-I00, by the Junta de Extremadura (Spain) and Fondo Europeo de Desarrollo Regional (FEDER, EU) through Grants No.\ GR21014 and No.\ IB20079, and by Comunidad de Madrid (Spain) under the Multiannual Agreement with UC3M in the line of Excellence of University Professors (EPUC3M23), in the context of the V Plan Regional de Investigaci\'on Cient\'{\i}fica e Innovaci\'on Tecnol\'ogica (PRICIT). B.\ G.\ Barreales was supported by Junta de Extremadura and Fondo Social Europeo (FSE, EU) through pre-doctoral grant PD18034. We have run our simulations in the computing facilities of the Instituto de Computaci\'{o}n Cient\'{\i}fica Avanzada de Extremadura (ICCAEx).
\end{acknowledgments}

\appendix

\section{Simulation details} 
\label{app:details}

The parameters employed in our simulations are listed in Table \ref{tab:details}. The maximum time $t_\mathrm{max}$ of the simulations has been chosen in such a way that approximately half of the runs survive and have not entered the absorbing state before this time. The lattice size is $L^d$. For reference, one run in three dimensions and $L=128$ takes approximately one month in our computer clusters (AMD processors). 

\begin{table}[b]
\caption{Parameter values for our numerical simulations.}
\label{tab:details}
\begin{ruledtabular}
\begin{tabular}{crrr}
 $d$ & \multicolumn{1}{c}{$L$} & \multicolumn{1}{c}{$t_{\text{max}}$} & \#runs \\
\hline
1 & 512 & 60 $\times 10^3$ & 2000  \\
  & 1024 & 185 $\times 10^3$ & 2000  \\
  & 2048 & 530 $\times 10^3$ & 2000 \\
  & 4096 & 1600 $\times 10^3$ & 2000  \\
  & 8192 & 4000 $\times 10^3$ & 2000 \\
\hline
2 & 128 & 20 $\times 10^3$ & 500 \\
  & 256 & 50 $\times 10^3$ & 500 \\
  & 512 & 200 $\times 10^3$ & 498 \\
\hline
3 & 16 & 1 $\times 10^3$  & 100 \\
  & 32 & 3.5 $\times 10^3$ & 100 \\
  & 64 & 15 $\times 10^3$ & 100  \\
  & 128 & 40 $\times 10^3$ & 20 \\
\end{tabular}
\end{ruledtabular}
\end{table}

\section{Front structure factor} 
\label{app:SF}

As mentioned in Sec.~\ref{sect:observables}, the front or height structure factor gives us a complementary perspective on the anomalous scaling, which is particularly useful in the context of crossover behavior \cite{Siegert1996} and/or large roughness exponent values \cite{Lopez1997,Lopez1997b,Ramasco2000,Cuerno2004}. Specifically, the front structure factor $S(\boldsymbol{k},t)$ is defined as \cite{Barabasi1995,Krug1997}
\begin{equation}
    S(\boldsymbol{k},t)=\langle |\mathcal{F}[h(\boldsymbol{x},t)]|^2 \rangle ,
\end{equation}
where $\mathcal F$ denotes the space Fourier transform and $\boldsymbol{k}$ is $d$-dimensional wave vector. For isotropic systems displaying intrinsic anomalous scaling as in our case, $S(\boldsymbol{k},t)$ behaves as \cite{Lopez1997}
\begin{equation}
S(k,t) = k^{-(2 \alpha +d)} s(k t^{1/z}) ,
\label{eq:Skanom}
\end{equation}
where $s(y) \propto y^{2(\alpha - \alpha_{\rm loc})}$ for $y\gg 1$, $s(y) \propto y^{2\alpha +d}$ for $y\ll 1$, and $k=|\boldsymbol{k}|$. Analogously to the case with the height-difference correlation function, Eq.\ \eqref{eq:Skanom} generalizes the FV Ansatz for the structure factor \cite{Barabasi1995,Krug1997}, which is retrieved for $\alpha_{\rm loc}=\alpha$. In case of intrinsic anomalous scaling, two main implications of Eq.\ \eqref{eq:Skanom} should be stressed: {\em (i)} for large $k\gg t^{-1/z}$, the scaling of the structure factor with $k$ reveals the local roughness exponent, namely, $S(k) \sim k^{-(2\alpha_{\rm loc}+d)}$; {\em (ii)} the $S(k,t)$ curves as functions of $k$ do not overlap for different times.
 
Both these features {\em (i)} and {\em (ii)} are indeed found in our numerical data for the CP. Figure \ref{fig:factorS} shows the structure factor versus the modulus of the wave vector for different times, for $d=1, 2$, and $3$. For each dimension, the time shift of the curves is clear from the graphs, while the scaling of the high $k$ data agrees well with the expected $k^{-(2\alpha_{\rm loc}+d)}$ law using the value of $\alpha_{\rm loc}$ computed in Sec.\ \ref{sect:c2}, which is shown as solid lines in the various figure panels. Hence, as expected, the scaling behavior in Fourier space is consistent with that already found in real space in the main text.
Note that in the 3D case, in which our results differ more from those reported in Ref.\ \cite{Dickman2000}, we have rescaled our $S(k,t)$ data (not shown), using both our values of $\alpha$, $\alpha_{\rm loc}$, and $z$, as well as the ones provided in that reference; collapse is achieved in both cases within error bars.

 \begin{figure*}[htb]
    \centering
    \includegraphics[width=0.33\textwidth]{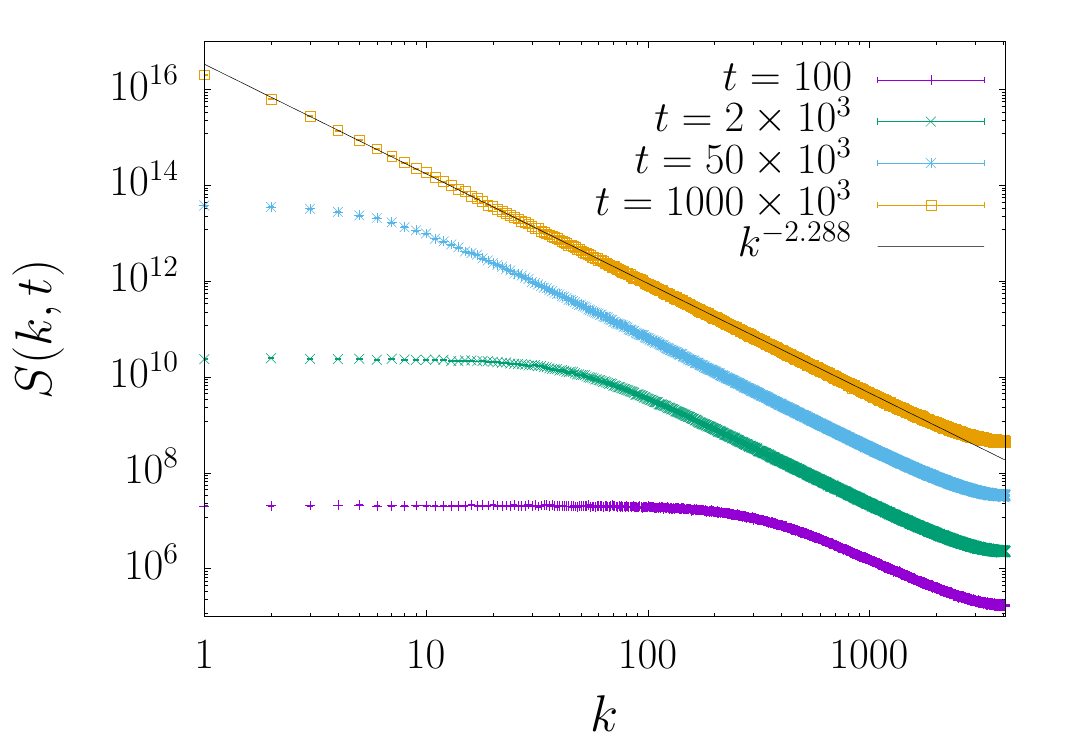}
    \includegraphics[width=0.33\textwidth]{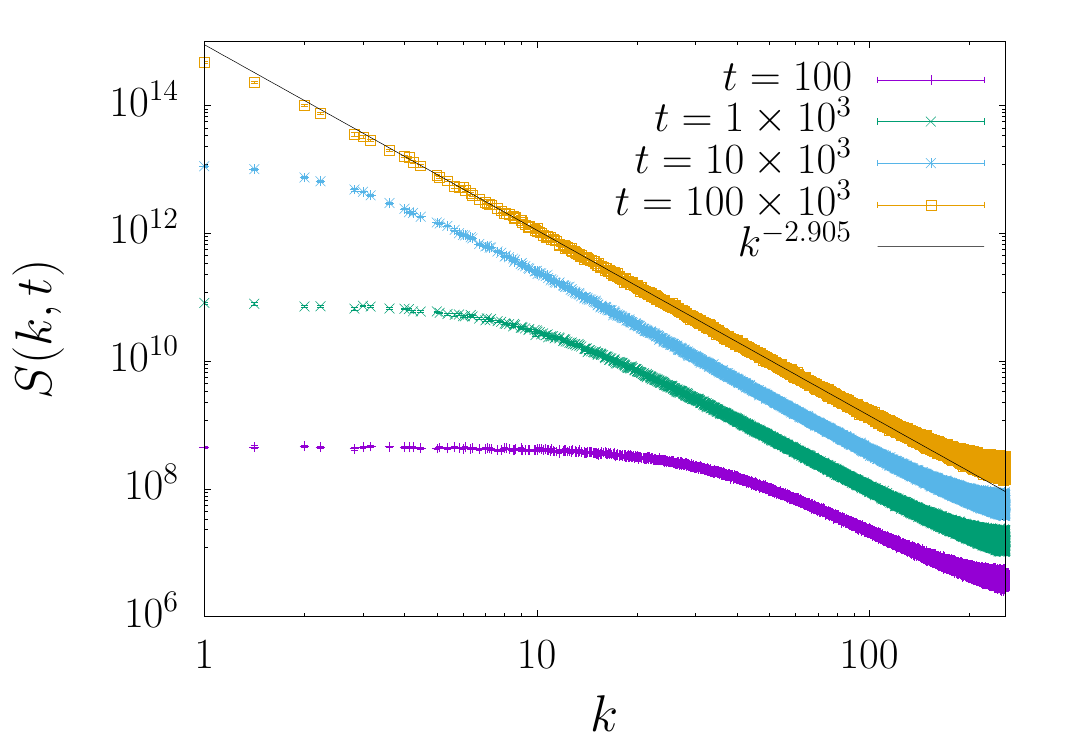}
    \includegraphics[width=0.33\textwidth]{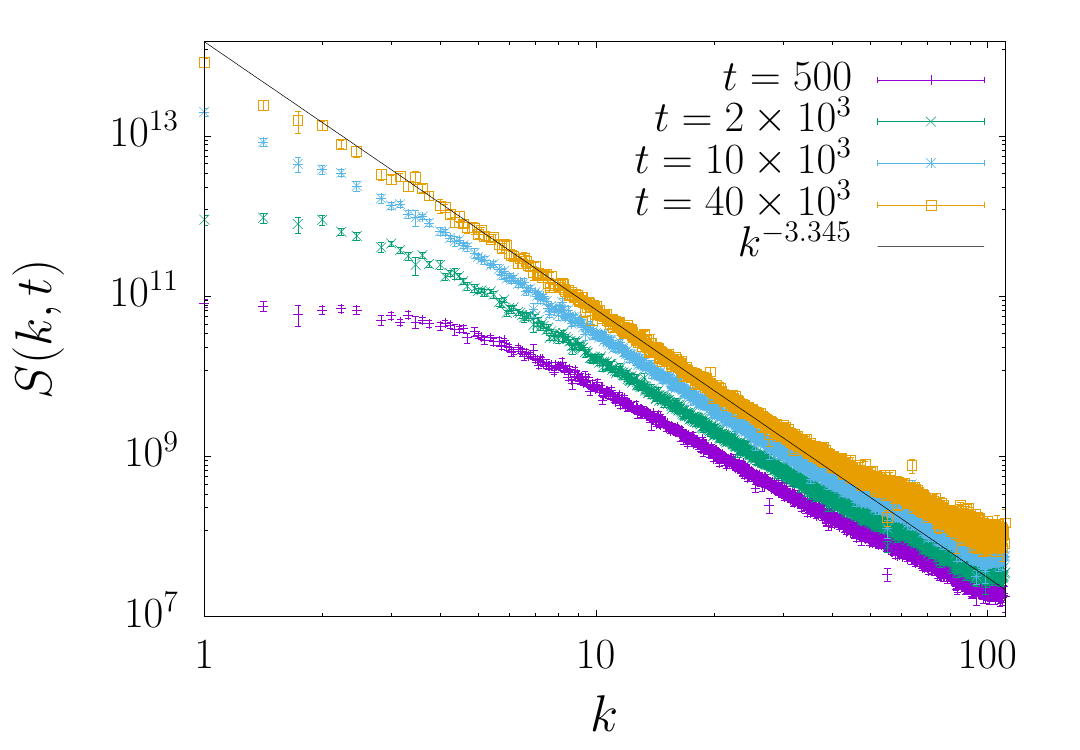}
    \caption{Structure factor $S(k,t)$ for one, two, and three dimensions and $L=8192, 512$, and $128$, respectively, at several times, as indicated in each legend. Solid lines correspond to $k^{-(2\alpha_{\rm loc}+d)}$, where $\alpha_{\rm loc}$ was computed from the scaling behavior found for $C_2(r,t)$, see Table \ref{tab:exponents}. }
    \label{fig:factorS}
\end{figure*}

\section{Mean squared height gradient} 
\label{app:gradient}
Anomalous scaling is related with a non-trivial time evolution for the slope field $\nabla h(\boldsymbol{x},t)$ \cite{Schroeder93,Krug1997}. Indeed, one can estimate the squared slope through the value of the height-difference correlation function evaluated at a distance of one lattice spacing $\Delta x$, so that $\langle (\nabla h)^2\rangle \approx (\Delta x)^2 C_2(\Delta x,t)$, which under the FV Ansatz becomes time-independent early-on in the time evolution \cite{Barabasi1995,Krug1997}. In contrast, in the presence of intrinsic anomalous scaling this quantity only saturates at steady state when $t=t_{\rm sat} \sim L^z$. In this case, assuming $\langle \overline{(\nabla h)^2}\rangle \sim t^{2\kappa}$, where $\kappa$ is an exponent characterizing the anomalous time increase of the average front slopes, the following scaling relation is expected to hold \cite{Lopez1999}:
\begin{equation}
    \alpha_{\rm loc}=\alpha - z\kappa .
    \label{eq:kappa}
\end{equation}
This scaling law was verified by the simulation results obtained in Ref.\ \cite{Dickman2000} and we consider it here in face of our numerical results.
To address it, we need to compute the mean squared height gradient $\overline{(\nabla h)^2 }$. In particular, we approximate the $j$th component of the $d$-dimensional gradient of $h(\boldsymbol{x})$ as
\begin{equation}
    \partial h/\partial x_j \approx  \dfrac{h(\boldsymbol{x}+(\Delta x) \boldsymbol{e}_j)-h(\boldsymbol{x})}{\Delta x} , \quad j=1,\ldots,d,
\end{equation}
where $\boldsymbol{e}_j$ is the $j$th vector of the canonical basis in $\mathbb{R}^d$ and $\Delta x=1$ in our lattice. 

In our simulations, the mean squared height gradient indeed increases as a power law for all values of $d$, see Fig.\ \ref{fig:grad}. The results of our fits lead to the values of the $\kappa$ exponent collected in Table \ref{tab:kappa}. Inserting our result for $\kappa$, $z$, and $\alpha$ in Eq.\ \eqref{eq:kappa}, we obtain a new estimate of $\alpha_{\rm loc}$, see Table \ref{tab:kappa}. These results are in agreement (at least for the largest system sizes) with the values of $\alpha_{\rm loc}$ directly obtained from the behavior of $C_2(r,t)$ in Sec.\ \ref{sect:c2}, recall Table \ref{tab:exponents}.

\begin{figure*}[htb]
    \centering
    \includegraphics[width=0.33\textwidth]{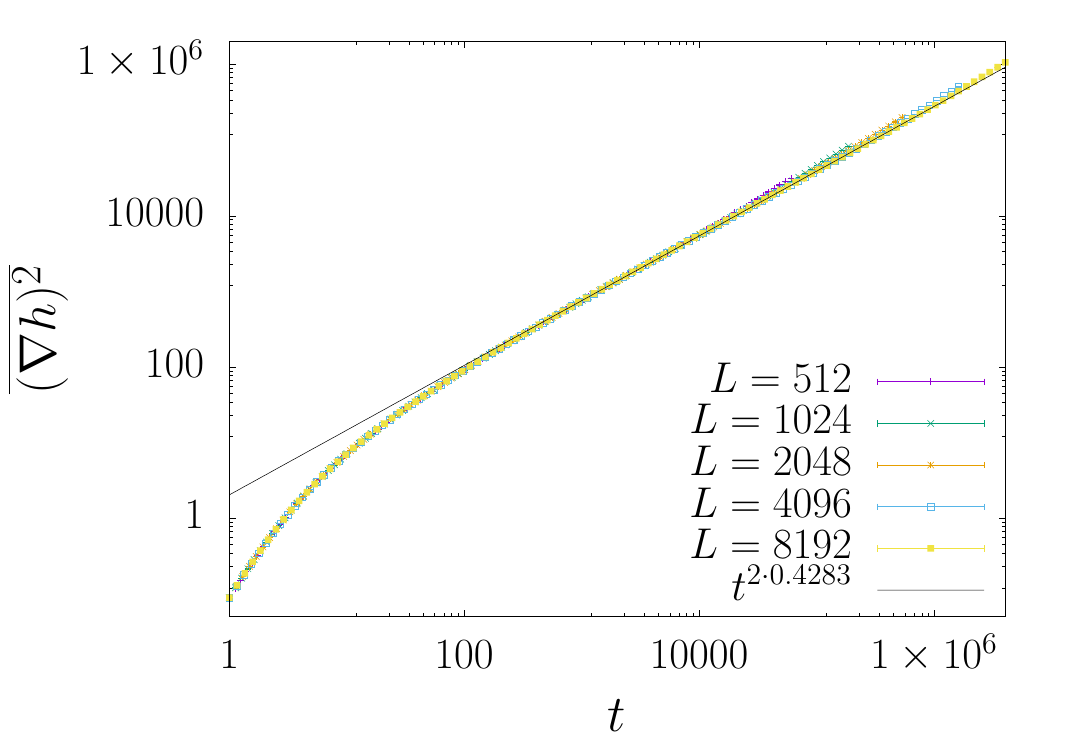}
    \includegraphics[width=0.33\textwidth]{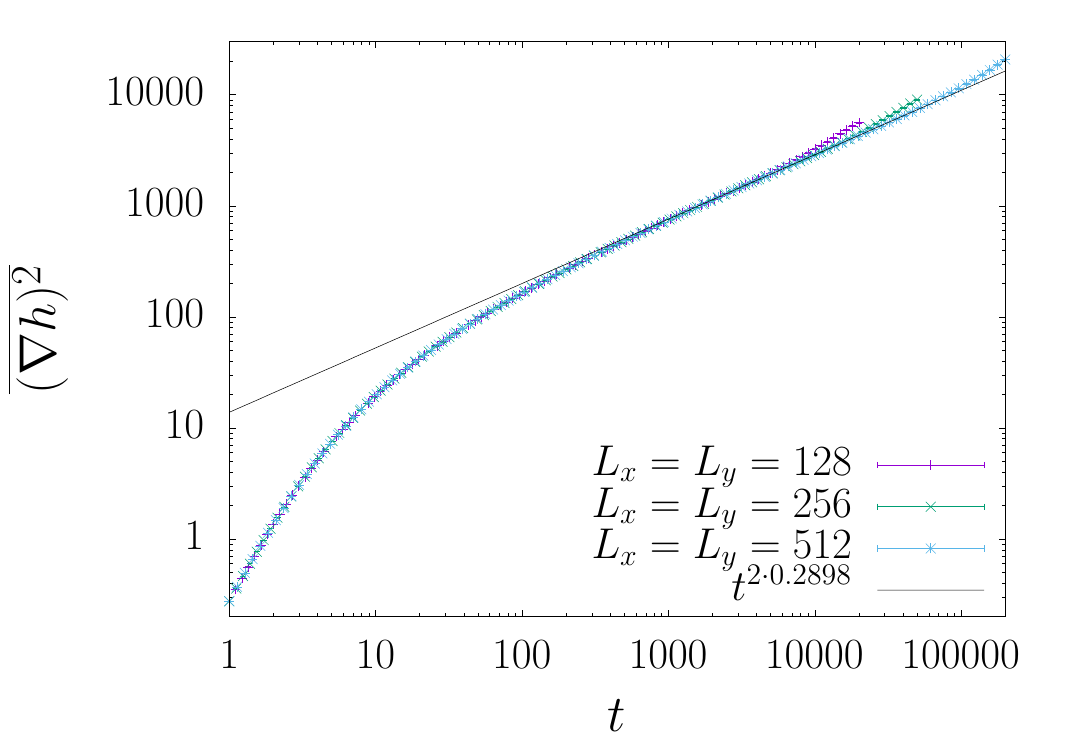}
    \includegraphics[width=0.33\textwidth]{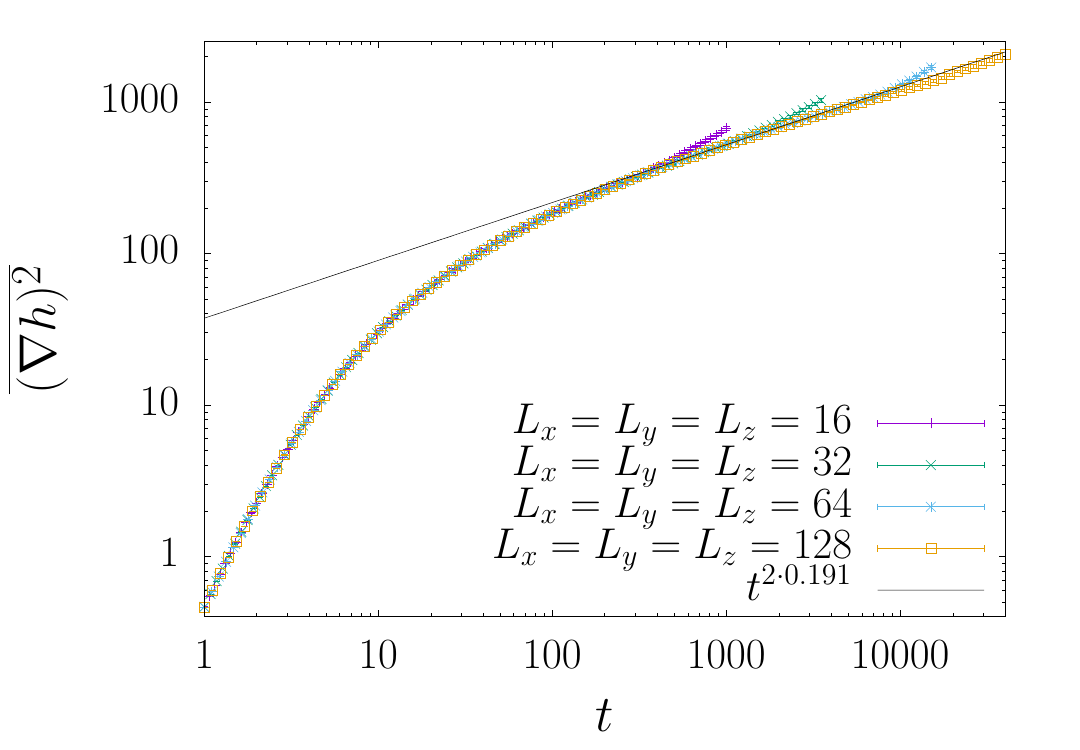}
    \caption{Mean squared height gradient for one, two, and three dimensions, left to right, and different sizes $L$, as indicated in the legend. Solid lines show the best fit to the data of the largest $L$.}
    \label{fig:grad}
\end{figure*}

\begin{table}[htb]
\caption{Exponents $\kappa$ and $\alpha_{\rm loc}$ as obtained from Eq.\ \eqref{eq:kappa}, for different dimensions and system sizes.}
\label{tab:kappa}
\begin{ruledtabular}
\begin{tabular}{lrll}
$d$ & $L$ & $\kappa$ & $\alpha_{\rm loc}$ \\ 
\colrule
1 & 512 & 0.4396(11) & 0.643(10) \\
  & 1024 & 0.4329(9) & 0.649(7) \\ 
  & 2048 & 0.4297(7) & 0.652(9) \\ 
  & 4096 & 0.4258(6) & 0.657(4) \\ 
  & 8192 &  0.4283(18) & 0.650(4) \\ 
\colrule
2 & 128 & 0.304(3) &  0.466(12) \\
  & 256 &  0.2955(13) & 0.458(7) \\ 
  & 512 &  0.2898(10)  & 0.458(7) \\ 
\colrule
3 & 64 & 0.2018(18) & 0.240(18)\\ 
  & 128 & 0.191(2) & 0.21(4)\\ 
\end{tabular}
\end{ruledtabular}
\end{table}


\begin{thebibliography}{60}%
\makeatletter
\providecommand \@ifxundefined [1]{%
 \@ifx{#1\undefined}
}%
\providecommand \@ifnum [1]{%
 \ifnum #1\expandafter \@firstoftwo
 \else \expandafter \@secondoftwo
 \fi
}%
\providecommand \@ifx [1]{%
 \ifx #1\expandafter \@firstoftwo
 \else \expandafter \@secondoftwo
 \fi
}%
\providecommand \natexlab [1]{#1}%
\providecommand \enquote  [1]{``#1''}%
\providecommand \bibnamefont  [1]{#1}%
\providecommand \bibfnamefont [1]{#1}%
\providecommand \citenamefont [1]{#1}%
\providecommand \href@noop [0]{\@secondoftwo}%
\providecommand \href [0]{\begingroup \@sanitize@url \@href}%
\providecommand \@href[1]{\@@startlink{#1}\@@href}%
\providecommand \@@href[1]{\endgroup#1\@@endlink}%
\providecommand \@sanitize@url [0]{\catcode `\\12\catcode `\$12\catcode
  `\&12\catcode `\#12\catcode `\^12\catcode `\_12\catcode `\%12\relax}%
\providecommand \@@startlink[1]{}%
\providecommand \@@endlink[0]{}%
\providecommand \url  [0]{\begingroup\@sanitize@url \@url }%
\providecommand \@url [1]{\endgroup\@href {#1}{\urlprefix }}%
\providecommand \urlprefix  [0]{URL }%
\providecommand \Eprint [0]{\href }%
\providecommand \doibase [0]{https://doi.org/}%
\providecommand \selectlanguage [0]{\@gobble}%
\providecommand \bibinfo  [0]{\@secondoftwo}%
\providecommand \bibfield  [0]{\@secondoftwo}%
\providecommand \translation [1]{[#1]}%
\providecommand \BibitemOpen [0]{}%
\providecommand \bibitemStop [0]{}%
\providecommand \bibitemNoStop [0]{.\EOS\space}%
\providecommand \EOS [0]{\spacefactor3000\relax}%
\providecommand \BibitemShut  [1]{\csname bibitem#1\endcsname}%
\let\auto@bib@innerbib\@empty
\bibitem [{\citenamefont {T{\"{a}}uber}(2014)}]{Tauber2014}%
  \BibitemOpen
  \bibfield  {author} {\bibinfo {author} {\bibfnamefont {U.~C.}\ \bibnamefont
  {T{\"{a}}uber}},\ }\href {https://doi.org/10.1017/CBO9781139046213} {\emph
  {\bibinfo {title} {Critical Dynamics: A Field Theory Approach to Equilibrium
  and Non-Equilibrium Scaling Behavior}}}\ (\bibinfo  {publisher} {Cambridge
  University Press},\ \bibinfo {address} {Cambridge, UK},\ \bibinfo {year}
  {2014})\BibitemShut {NoStop}%
\bibitem [{\citenamefont {Barab{\'{a}}si}\ and\ \citenamefont
  {Stanley}(1995)}]{Barabasi1995}%
  \BibitemOpen
  \bibfield  {author} {\bibinfo {author} {\bibfnamefont {A.-L.}\ \bibnamefont
  {Barab{\'{a}}si}}\ and\ \bibinfo {author} {\bibfnamefont {H.~E.}\
  \bibnamefont {Stanley}},\ }\href {https://doi.org/10.1017/cbo9780511599798}
  {\emph {\bibinfo {title} {{Fractal Concepts in Surface Growth}}}}\ (\bibinfo
  {publisher} {Cambridge University Press},\ \bibinfo {address} {Cambridge,
  UK},\ \bibinfo {year} {1995})\BibitemShut {NoStop}%
\bibitem [{\citenamefont {Krug}(1997)}]{Krug1997}%
  \BibitemOpen
  \bibfield  {author} {\bibinfo {author} {\bibfnamefont {J.}~\bibnamefont
  {Krug}},\ }\href {https://doi.org/10.1080/00018739700101498} {\bibfield
  {journal} {\bibinfo  {journal} {Adv. Phys.}\ }\textbf {\bibinfo {volume}
  {46}},\ \bibinfo {pages} {139} (\bibinfo {year} {1997})}\BibitemShut
  {NoStop}%
\bibitem [{\citenamefont {Kriecherbauer}\ and\ \citenamefont
  {Krug}(2010)}]{kriecherbauer2010}%
  \BibitemOpen
  \bibfield  {author} {\bibinfo {author} {\bibfnamefont {T.}~\bibnamefont
  {Kriecherbauer}}\ and\ \bibinfo {author} {\bibfnamefont {J.}~\bibnamefont
  {Krug}},\ }\href {https://doi.org/10.1088/1751-8113/43/40/403001} {\bibfield
  {journal} {\bibinfo  {journal} {J. Phys. A: Math. Theor.}\ }\textbf {\bibinfo
  {volume} {43}},\ \bibinfo {pages} {403001} (\bibinfo {year}
  {2010})}\BibitemShut {NoStop}%
\bibitem [{\citenamefont {Halpin-Healy}\ and\ \citenamefont
  {Takeuchi}(2015)}]{Halpin-Healy2015}%
  \BibitemOpen
  \bibfield  {author} {\bibinfo {author} {\bibfnamefont {T.}~\bibnamefont
  {Halpin-Healy}}\ and\ \bibinfo {author} {\bibfnamefont {K.~A.}\ \bibnamefont
  {Takeuchi}},\ }\href {https://doi.org/10.1007/s10955-015-1282-1} {\bibfield
  {journal} {\bibinfo  {journal} {J. Stat. Phys.}\ }\textbf {\bibinfo {volume}
  {160}},\ \bibinfo {pages} {794} (\bibinfo {year} {2015})}\BibitemShut
  {NoStop}%
\bibitem [{\citenamefont {Takeuchi}(2018)}]{Takeuchi2018}%
  \BibitemOpen
  \bibfield  {author} {\bibinfo {author} {\bibfnamefont {K.~A.}\ \bibnamefont
  {Takeuchi}},\ }\href {https://doi.org/10.1016/j.physa.2018.03.009} {\bibfield
   {journal} {\bibinfo  {journal} {Physica A}\ }\textbf {\bibinfo {volume}
  {504}},\ \bibinfo {pages} {77} (\bibinfo {year} {2018})}\BibitemShut
  {NoStop}%
\bibitem [{\citenamefont {Carrasco}\ and\ \citenamefont
  {Oliveira}(2019)}]{Carrasco2019}%
  \BibitemOpen
  \bibfield  {author} {\bibinfo {author} {\bibfnamefont {I.~S.}\ \bibnamefont
  {Carrasco}}\ and\ \bibinfo {author} {\bibfnamefont {T.~J.}\ \bibnamefont
  {Oliveira}},\ }\href {https://doi.org/10.1103/PhysRevE.100.042107} {\bibfield
   {journal} {\bibinfo  {journal} {Phys. Rev. E}\ }\textbf {\bibinfo {volume}
  {100}},\ \bibinfo {pages} {042107} (\bibinfo {year} {2019})}\BibitemShut
  {NoStop}%
\bibitem [{\citenamefont {Carrasco}\ and\ \citenamefont
  {Oliveira}(2016)}]{Carrasco2016}%
  \BibitemOpen
  \bibfield  {author} {\bibinfo {author} {\bibfnamefont {I.~S.~S.}\
  \bibnamefont {Carrasco}}\ and\ \bibinfo {author} {\bibfnamefont {T.~J.}\
  \bibnamefont {Oliveira}},\ }\href
  {https://doi.org/10.1103/PhysRevE.94.050801} {\bibfield  {journal} {\bibinfo
  {journal} {Phys. Rev. E}\ }\textbf {\bibinfo {volume} {94}},\ \bibinfo
  {pages} {050801(R)} (\bibinfo {year} {2016})}\BibitemShut {NoStop}%
\bibitem [{\citenamefont {Rodr{\'{i}}guez-Fern{\'{a}}ndez}\ and\ \citenamefont
  {Cuerno}(2020)}]{Rodriguez-Fernandez2020}%
  \BibitemOpen
  \bibfield  {author} {\bibinfo {author} {\bibfnamefont {E.}~\bibnamefont
  {Rodr{\'{i}}guez-Fern{\'{a}}ndez}}\ and\ \bibinfo {author} {\bibfnamefont
  {R.}~\bibnamefont {Cuerno}},\ }\href
  {https://doi.org/10.1103/PhysRevE.101.052126} {\bibfield  {journal} {\bibinfo
   {journal} {Phys. Rev. E}\ }\textbf {\bibinfo {volume} {101}},\ \bibinfo
  {pages} {052126} (\bibinfo {year} {2020})}\BibitemShut {NoStop}%
\bibitem [{\citenamefont {Halpin-Healy}(2012)}]{Halpin-Healy2012}%
  \BibitemOpen
  \bibfield  {author} {\bibinfo {author} {\bibfnamefont {T.}~\bibnamefont
  {Halpin-Healy}},\ }\href {https://doi.org/10.1103/PhysRevLett.109.170602}
  {\bibfield  {journal} {\bibinfo  {journal} {Phys. Rev. Lett.}\ }\textbf
  {\bibinfo {volume} {109}},\ \bibinfo {pages} {170602} (\bibinfo {year}
  {2012})}\BibitemShut {NoStop}%
\bibitem [{\citenamefont {Halpin-Healy}(2013)}]{Halpin-Healy2013}%
  \BibitemOpen
  \bibfield  {author} {\bibinfo {author} {\bibfnamefont {T.}~\bibnamefont
  {Halpin-Healy}},\ }\href {https://doi.org/10.1103/PhysRevE.88.042118}
  {\bibfield  {journal} {\bibinfo  {journal} {Phys. Rev. E}\ }\textbf {\bibinfo
  {volume} {88}},\ \bibinfo {pages} {042118} (\bibinfo {year}
  {2013})}\BibitemShut {NoStop}%
\bibitem [{\citenamefont {Oliveira}\ \emph {et~al.}(2012)\citenamefont
  {Oliveira}, \citenamefont {Ferreira},\ and\ \citenamefont
  {Alves}}]{Oliveira2012}%
  \BibitemOpen
  \bibfield  {author} {\bibinfo {author} {\bibfnamefont {T.~J.}\ \bibnamefont
  {Oliveira}}, \bibinfo {author} {\bibfnamefont {S.~C.}\ \bibnamefont
  {Ferreira}},\ and\ \bibinfo {author} {\bibfnamefont {S.~G.}\ \bibnamefont
  {Alves}},\ }\href {https://doi.org/10.1103/PhysRevE.85.010601} {\bibfield
  {journal} {\bibinfo  {journal} {Phys. Rev. E}\ }\textbf {\bibinfo {volume}
  {85}},\ \bibinfo {pages} {010601(R)} (\bibinfo {year} {2012})}\BibitemShut
  {NoStop}%
\bibitem [{\citenamefont {Oliveira}\ \emph {et~al.}(2013)\citenamefont
  {Oliveira}, \citenamefont {Alves},\ and\ \citenamefont
  {Ferreira}}]{Oliveira2013}%
  \BibitemOpen
  \bibfield  {author} {\bibinfo {author} {\bibfnamefont {T.~J.}\ \bibnamefont
  {Oliveira}}, \bibinfo {author} {\bibfnamefont {S.~G.}\ \bibnamefont
  {Alves}},\ and\ \bibinfo {author} {\bibfnamefont {S.~C.}\ \bibnamefont
  {Ferreira}},\ }\href {https://doi.org/10.1103/PhysRevE.87.040102} {\bibfield
  {journal} {\bibinfo  {journal} {Phys. Rev. E}\ }\textbf {\bibinfo {volume}
  {87}},\ \bibinfo {pages} {040102} (\bibinfo {year} {2013})}\BibitemShut
  {NoStop}%
\bibitem [{\citenamefont {Halpin-Healy}\ and\ \citenamefont
  {Palasantzas}(2014)}]{Halpin-Healy2014}%
  \BibitemOpen
  \bibfield  {author} {\bibinfo {author} {\bibfnamefont {T.}~\bibnamefont
  {Halpin-Healy}}\ and\ \bibinfo {author} {\bibfnamefont {G.}~\bibnamefont
  {Palasantzas}},\ }\href {https://doi.org/10.1209/0295-5075/105/50001}
  {\bibfield  {journal} {\bibinfo  {journal} {EPL}\ }\textbf {\bibinfo {volume}
  {105}},\ \bibinfo {pages} {50001} (\bibinfo {year} {2014})}\BibitemShut
  {NoStop}%
\bibitem [{\citenamefont {Almeida}\ \emph {et~al.}(2014)\citenamefont
  {Almeida}, \citenamefont {Ferreira}, \citenamefont {Oliveira},\ and\
  \citenamefont {Reis}}]{Almeida2014}%
  \BibitemOpen
  \bibfield  {author} {\bibinfo {author} {\bibfnamefont {R.~A.~L.}\
  \bibnamefont {Almeida}}, \bibinfo {author} {\bibfnamefont {S.~O.}\
  \bibnamefont {Ferreira}}, \bibinfo {author} {\bibfnamefont {T.~J.}\
  \bibnamefont {Oliveira}},\ and\ \bibinfo {author} {\bibfnamefont {F.~D. A.
  A.~a.}\ \bibnamefont {Reis}},\ }\href
  {https://doi.org/10.1103/PhysRevB.89.045309} {\bibfield  {journal} {\bibinfo
  {journal} {Phys. Rev. B}\ }\textbf {\bibinfo {volume} {89}},\ \bibinfo
  {pages} {045309} (\bibinfo {year} {2014})}\BibitemShut {NoStop}%
\bibitem [{\citenamefont {Henkel}\ \emph {et~al.}(2008)\citenamefont {Henkel},
  \citenamefont {Hinrichsen},\ and\ \citenamefont {Lübeck}}]{Henkel2008}%
  \BibitemOpen
  \bibfield  {author} {\bibinfo {author} {\bibfnamefont {M.}~\bibnamefont
  {Henkel}}, \bibinfo {author} {\bibfnamefont {H.}~\bibnamefont {Hinrichsen}},\
  and\ \bibinfo {author} {\bibfnamefont {S.}~\bibnamefont {Lübeck}},\ }\href
  {https://doi.org/10.1007/978-1-4020-8765-3} {\emph {\bibinfo {title}
  {Non-Equilibrium Phase Transitions, Vol. 1}}}\ (\bibinfo  {publisher}
  {Springer},\ \bibinfo {address} {Dordrecht},\ \bibinfo {year}
  {2008})\BibitemShut {NoStop}%
\bibitem [{\citenamefont {Rodr{\'{i}}guez-Fern{\'{a}}ndez}\ and\ \citenamefont
  {Cuerno}(2021)}]{Rodriguez-Fernandez2021}%
  \BibitemOpen
  \bibfield  {author} {\bibinfo {author} {\bibfnamefont {E.}~\bibnamefont
  {Rodr{\'{i}}guez-Fern{\'{a}}ndez}}\ and\ \bibinfo {author} {\bibfnamefont
  {R.}~\bibnamefont {Cuerno}},\ }\href
  {https://doi.org/10.1103/PhysRevResearch.3.L012020} {\bibfield  {journal}
  {\bibinfo  {journal} {Phys. Rev. Research}\ }\textbf {\bibinfo {volume}
  {3}},\ \bibinfo {pages} {L012020} (\bibinfo {year} {2021})}\BibitemShut
  {NoStop}%
\bibitem [{\citenamefont {Marcos}\ \emph {et~al.}(2022)\citenamefont {Marcos},
  \citenamefont {Rodr{\'{i}}guez-L{\'{o}}pez}, \citenamefont {Mel{\'{e}}ndez},
  \citenamefont {Cuerno},\ and\ \citenamefont {Ruiz-Lorenzo}}]{Marcos2022}%
  \BibitemOpen
  \bibfield  {author} {\bibinfo {author} {\bibfnamefont {J.~M.}\ \bibnamefont
  {Marcos}}, \bibinfo {author} {\bibfnamefont {P.}~\bibnamefont
  {Rodr{\'{i}}guez-L{\'{o}}pez}}, \bibinfo {author} {\bibfnamefont {J.~J.}\
  \bibnamefont {Mel{\'{e}}ndez}}, \bibinfo {author} {\bibfnamefont
  {R.}~\bibnamefont {Cuerno}},\ and\ \bibinfo {author} {\bibfnamefont {J.~J.}\
  \bibnamefont {Ruiz-Lorenzo}},\ }\href
  {https://doi.org/10.1103/physreve.105.054801} {\bibfield  {journal} {\bibinfo
   {journal} {Phys. Rev. E}\ }\textbf {\bibinfo {volume} {105}},\ \bibinfo
  {pages} {1} (\bibinfo {year} {2022})}\BibitemShut {NoStop}%
\bibitem [{\citenamefont {Fortin}\ and\ \citenamefont
  {Clusel}(2015)}]{Fortin2015}%
  \BibitemOpen
  \bibfield  {author} {\bibinfo {author} {\bibfnamefont {J.-Y.}\ \bibnamefont
  {Fortin}}\ and\ \bibinfo {author} {\bibfnamefont {M.}~\bibnamefont
  {Clusel}},\ }\href {https://doi.org/10.1088/1751-8113/48/18/183001}
  {\bibfield  {journal} {\bibinfo  {journal} {J. Phys. A: Math. Theor.}\
  }\textbf {\bibinfo {volume} {48}},\ \bibinfo {pages} {183001} (\bibinfo
  {year} {2015})}\BibitemShut {NoStop}%
\bibitem [{\citenamefont {Alves}\ \emph {et~al.}(2014)\citenamefont {Alves},
  \citenamefont {Oliveira},\ and\ \citenamefont {Ferreira}}]{Alves2014}%
  \BibitemOpen
  \bibfield  {author} {\bibinfo {author} {\bibfnamefont {S.~G.}\ \bibnamefont
  {Alves}}, \bibinfo {author} {\bibfnamefont {T.~J.}\ \bibnamefont
  {Oliveira}},\ and\ \bibinfo {author} {\bibfnamefont {S.~C.}\ \bibnamefont
  {Ferreira}},\ }\href {https://doi.org/10.1103/PhysRevE.90.020103} {\bibfield
  {journal} {\bibinfo  {journal} {Phys. Rev. E}\ }\textbf {\bibinfo {volume}
  {90}},\ \bibinfo {pages} {020103(R)} (\bibinfo {year} {2014})}\BibitemShut
  {NoStop}%
\bibitem [{\citenamefont {Schroeder}\ and\ \citenamefont
  {Siegert}(1993)}]{Schroeder93}%
  \BibitemOpen
  \bibfield  {author} {\bibinfo {author} {\bibfnamefont {M.}~\bibnamefont
  {Schroeder}}\ and\ \bibinfo {author} {\bibfnamefont {M.}~\bibnamefont
  {Siegert}},\ }\href {https://doi.org/10.1209/0295-5075/24/7/010} {\bibfield
  {journal} {\bibinfo  {journal} {Europhys. Lett.}\ }\textbf {\bibinfo {volume}
  {24}},\ \bibinfo {pages} {563} (\bibinfo {year} {1993})}\BibitemShut
  {NoStop}%
\bibitem [{\citenamefont {Das~Sarma}\ \emph {et~al.}(1994)\citenamefont
  {Das~Sarma}, \citenamefont {Ghaisas},\ and\ \citenamefont
  {Kim}}]{Dassarma94}%
  \BibitemOpen
  \bibfield  {author} {\bibinfo {author} {\bibfnamefont {S.}~\bibnamefont
  {Das~Sarma}}, \bibinfo {author} {\bibfnamefont {S.~V.}\ \bibnamefont
  {Ghaisas}},\ and\ \bibinfo {author} {\bibfnamefont {J.~M.}\ \bibnamefont
  {Kim}},\ }\href {https://doi.org/10.1103/PhysRevE.49.122} {\bibfield
  {journal} {\bibinfo  {journal} {Phys. Rev. E}\ }\textbf {\bibinfo {volume}
  {49}},\ \bibinfo {pages} {122} (\bibinfo {year} {1994})}\BibitemShut
  {NoStop}%
\bibitem [{\citenamefont {L\'opez}\ \emph {et~al.}(1997)\citenamefont
  {L\'opez}, \citenamefont {Rodr\'{\i}guez},\ and\ \citenamefont
  {Cuerno}}]{Lopez1997b}%
  \BibitemOpen
  \bibfield  {author} {\bibinfo {author} {\bibfnamefont {J.~M.}\ \bibnamefont
  {L\'opez}}, \bibinfo {author} {\bibfnamefont {M.~A.}\ \bibnamefont
  {Rodr\'{\i}guez}},\ and\ \bibinfo {author} {\bibfnamefont {R.}~\bibnamefont
  {Cuerno}},\ }\href {https://doi.org/10.1103/PhysRevE.56.3993} {\bibfield
  {journal} {\bibinfo  {journal} {Phys. Rev. E}\ }\textbf {\bibinfo {volume}
  {56}},\ \bibinfo {pages} {3993} (\bibinfo {year} {1997})}\BibitemShut
  {NoStop}%
\bibitem [{\citenamefont {Ramasco}\ \emph {et~al.}(2000)\citenamefont
  {Ramasco}, \citenamefont {Lopez},\ and\ \citenamefont
  {Rodriguez}}]{Ramasco2000}%
  \BibitemOpen
  \bibfield  {author} {\bibinfo {author} {\bibfnamefont {J.~J.}\ \bibnamefont
  {Ramasco}}, \bibinfo {author} {\bibfnamefont {J.~M.}\ \bibnamefont {Lopez}},\
  and\ \bibinfo {author} {\bibfnamefont {M.~A.}\ \bibnamefont {Rodriguez}},\
  }\href {https://doi.org/10.1103/PhysRevLett.84.2199} {\bibfield  {journal}
  {\bibinfo  {journal} {Phys. Rev. Lett.}\ }\textbf {\bibinfo {volume} {84}},\
  \bibinfo {pages} {2199} (\bibinfo {year} {2000})}\BibitemShut {NoStop}%
\bibitem [{\citenamefont {Cuerno}\ and\ \citenamefont
  {V\'{a}zquez}(2004)}]{Cuerno2004}%
  \BibitemOpen
  \bibfield  {author} {\bibinfo {author} {\bibfnamefont {R.}~\bibnamefont
  {Cuerno}}\ and\ \bibinfo {author} {\bibfnamefont {L.}~\bibnamefont
  {V\'{a}zquez}},\ }in\ \href {https://doi.org/10.48550/arXiv.cond-mat/0402630}
  {\emph {\bibinfo {booktitle} {Advances in Condensed Matter and Statistical
  Physics}}},\ \bibinfo {editor} {edited by\ \bibinfo {editor} {\bibfnamefont
  {E.}~\bibnamefont {Korutcheva}}\ and\ \bibinfo {editor} {\bibfnamefont
  {R.}~\bibnamefont {Cuerno}}}\ (\bibinfo  {publisher} {Nova Science
  Publishers},\ \bibinfo {address} {New York},\ \bibinfo {year}
  {2004})\BibitemShut {NoStop}%
\bibitem [{\citenamefont {Rodriguez-Fernandez}\ \emph
  {et~al.}(2022)\citenamefont {Rodriguez-Fernandez}, \citenamefont {Santalla},
  \citenamefont {Castro},\ and\ \citenamefont
  {Cuerno}}]{Rodriguez-Fernandez2022}%
  \BibitemOpen
  \bibfield  {author} {\bibinfo {author} {\bibfnamefont {E.}~\bibnamefont
  {Rodriguez-Fernandez}}, \bibinfo {author} {\bibfnamefont {S.~N.}\
  \bibnamefont {Santalla}}, \bibinfo {author} {\bibfnamefont {M.}~\bibnamefont
  {Castro}},\ and\ \bibinfo {author} {\bibfnamefont {R.}~\bibnamefont
  {Cuerno}},\ }\href {https://doi.org/10.1103/PhysRevE.106.024802} {\bibfield
  {journal} {\bibinfo  {journal} {Phys. Rev. E}\ }\textbf {\bibinfo {volume}
  {106}},\ \bibinfo {pages} {024802} (\bibinfo {year} {2022})}\BibitemShut
  {NoStop}%
\bibitem [{\citenamefont {Guti\'errez}\ and\ \citenamefont
  {Cuerno}(2023)}]{Gutierrez2023}%
  \BibitemOpen
  \bibfield  {author} {\bibinfo {author} {\bibfnamefont {R.}~\bibnamefont
  {Guti\'errez}}\ and\ \bibinfo {author} {\bibfnamefont {R.}~\bibnamefont
  {Cuerno}},\ }\href {https://doi.org/10.1103/PhysRevResearch.5.023047}
  {\bibfield  {journal} {\bibinfo  {journal} {Phys. Rev. Research}\ }\textbf
  {\bibinfo {volume} {5}},\ \bibinfo {pages} {023047} (\bibinfo {year}
  {2023})}\BibitemShut {NoStop}%
\bibitem [{\citenamefont {L{\'{o}}pez}\ \emph {et~al.}(2005)\citenamefont
  {L{\'{o}}pez}, \citenamefont {Castro},\ and\ \citenamefont
  {Gallego}}]{Lopez2005}%
  \BibitemOpen
  \bibfield  {author} {\bibinfo {author} {\bibfnamefont {J.}~\bibnamefont
  {L{\'{o}}pez}}, \bibinfo {author} {\bibfnamefont {M.}~\bibnamefont
  {Castro}},\ and\ \bibinfo {author} {\bibfnamefont {R.}~\bibnamefont
  {Gallego}},\ }\href {https://doi.org/10.1103/PhysRevLett.94.166103}
  {\bibfield  {journal} {\bibinfo  {journal} {Phys. Rev. Lett.}\ }\textbf
  {\bibinfo {volume} {94}},\ \bibinfo {pages} {166103} (\bibinfo {year}
  {2005})}\BibitemShut {NoStop}%
\bibitem [{\citenamefont {Liggett}(1985)}]{Liggett1985}%
  \BibitemOpen
  \bibfield  {author} {\bibinfo {author} {\bibfnamefont {T.~M.}\ \bibnamefont
  {Liggett}},\ }\href@noop {} {\emph {\bibinfo {title} {Interacting Particle
  Systems}}}\ (\bibinfo  {publisher} {Springer-Verlag},\ \bibinfo {address}
  {New York},\ \bibinfo {year} {1985})\BibitemShut {NoStop}%
\bibitem [{\citenamefont {{\'{O}}dor}(2004)}]{Odor2004}%
  \BibitemOpen
  \bibfield  {author} {\bibinfo {author} {\bibfnamefont {G.}~\bibnamefont
  {{\'{O}}dor}},\ }\href {https://doi.org/10.1103/RevModPhys.76.663} {\bibfield
   {journal} {\bibinfo  {journal} {Rev. Mod. Phys.}\ }\textbf {\bibinfo
  {volume} {76}},\ \bibinfo {pages} {663} (\bibinfo {year} {2004})}\BibitemShut
  {NoStop}%
\bibitem [{\citenamefont {Dickman}\ and\ \citenamefont
  {Muñoz}(2000)}]{Dickman2000}%
  \BibitemOpen
  \bibfield  {author} {\bibinfo {author} {\bibfnamefont {R.}~\bibnamefont
  {Dickman}}\ and\ \bibinfo {author} {\bibfnamefont {M.~A.}\ \bibnamefont
  {Muñoz}},\ }\href {https://doi.org/10.1103/PhysRevE.62.7632} {\bibfield
  {journal} {\bibinfo  {journal} {Phys. Rev. E}\ }\textbf {\bibinfo {volume}
  {62}},\ \bibinfo {pages} {7632} (\bibinfo {year} {2000})}\BibitemShut
  {NoStop}%
\bibitem [{\citenamefont {Szendro}\ \emph {et~al.}(2007)\citenamefont
  {Szendro}, \citenamefont {L\'opez},\ and\ \citenamefont
  {Rodr\'{\i}guez}}]{Szendro2007}%
  \BibitemOpen
  \bibfield  {author} {\bibinfo {author} {\bibfnamefont {I.~G.}\ \bibnamefont
  {Szendro}}, \bibinfo {author} {\bibfnamefont {J.~M.}\ \bibnamefont
  {L\'opez}},\ and\ \bibinfo {author} {\bibfnamefont {M.~A.}\ \bibnamefont
  {Rodr\'{\i}guez}},\ }\href {https://doi.org/10.1103/PhysRevE.76.011603}
  {\bibfield  {journal} {\bibinfo  {journal} {Phys. Rev. E}\ }\textbf {\bibinfo
  {volume} {76}},\ \bibinfo {pages} {011603} (\bibinfo {year}
  {2007})}\BibitemShut {NoStop}%
\bibitem [{\citenamefont {Song}\ and\ \citenamefont {Xia}(2021)}]{Song2021}%
  \BibitemOpen
  \bibfield  {author} {\bibinfo {author} {\bibfnamefont {T.}~\bibnamefont
  {Song}}\ and\ \bibinfo {author} {\bibfnamefont {H.}~\bibnamefont {Xia}},\
  }\href {https://doi.org/10.1088/1742-5468/ac06c3} {\bibfield  {journal}
  {\bibinfo  {journal} {J. Stat. Mech.}\ }\textbf {\bibinfo {volume} {2021}},\
  \bibinfo {pages} {073203} (\bibinfo {year} {2021})}\BibitemShut {NoStop}%
\bibitem [{\citenamefont {Wiese}(2022)}]{Wiese2022}%
  \BibitemOpen
  \bibfield  {author} {\bibinfo {author} {\bibfnamefont {K.~J.}\ \bibnamefont
  {Wiese}},\ }\href {https://doi.org/10.1088/1361-6633/ac4648} {\bibfield
  {journal} {\bibinfo  {journal} {Rep. Prog. Phys.}\ }\textbf {\bibinfo
  {volume} {85}},\ \bibinfo {pages} {086502} (\bibinfo {year}
  {2022})}\BibitemShut {NoStop}%
\bibitem [{\citenamefont {Barreales}\ \emph {et~al.}(2022)\citenamefont
  {Barreales}, \citenamefont {Mel{\'{e}}ndez}, \citenamefont {Cuerno},\ and\
  \citenamefont {Ruiz-Lorenzo}}]{Barreales2022}%
  \BibitemOpen
  \bibfield  {author} {\bibinfo {author} {\bibfnamefont {B.~G.}\ \bibnamefont
  {Barreales}}, \bibinfo {author} {\bibfnamefont {J.~J.}\ \bibnamefont
  {Mel{\'{e}}ndez}}, \bibinfo {author} {\bibfnamefont {R.}~\bibnamefont
  {Cuerno}},\ and\ \bibinfo {author} {\bibfnamefont {J.~J.}\ \bibnamefont
  {Ruiz-Lorenzo}},\ }\href {https://doi.org/10.1103/PhysRevE.106.044801}
  {\bibfield  {journal} {\bibinfo  {journal} {Phys. Rev. E}\ }\textbf {\bibinfo
  {volume} {106}},\ \bibinfo {pages} {044801} (\bibinfo {year}
  {2022})}\BibitemShut {NoStop}%
\bibitem [{\citenamefont {Rupp}\ \emph {et~al.}(2003)\citenamefont {Rupp},
  \citenamefont {Richter},\ and\ \citenamefont {Rehberg}}]{Rupp2003}%
  \BibitemOpen
  \bibfield  {author} {\bibinfo {author} {\bibfnamefont {P.}~\bibnamefont
  {Rupp}}, \bibinfo {author} {\bibfnamefont {R.}~\bibnamefont {Richter}},\ and\
  \bibinfo {author} {\bibfnamefont {I.}~\bibnamefont {Rehberg}},\ }\href
  {https://doi.org/10.1103/PhysRevE.67.036209} {\bibfield  {journal} {\bibinfo
  {journal} {Phys. Rev. E}\ }\textbf {\bibinfo {volume} {67}},\ \bibinfo
  {pages} {7} (\bibinfo {year} {2003})}\BibitemShut {NoStop}%
\bibitem [{\citenamefont {Takeuchi}\ \emph {et~al.}(2007)\citenamefont
  {Takeuchi}, \citenamefont {Kuroda}, \citenamefont {Chaté},\ and\
  \citenamefont {Sano}}]{Takeuchi2007}%
  \BibitemOpen
  \bibfield  {author} {\bibinfo {author} {\bibfnamefont {K.~A.}\ \bibnamefont
  {Takeuchi}}, \bibinfo {author} {\bibfnamefont {M.}~\bibnamefont {Kuroda}},
  \bibinfo {author} {\bibfnamefont {H.}~\bibnamefont {Chaté}},\ and\ \bibinfo
  {author} {\bibfnamefont {M.}~\bibnamefont {Sano}},\ }\bibfield  {journal}
  {\bibinfo  {journal} {Phys. Rev. Lett.}\ }\textbf {\bibinfo {volume} {99}},\
  \href {https://doi.org/10.1103/PhysRevLett.99.234503}
  {10.1103/PhysRevLett.99.234503} (\bibinfo {year} {2007})\BibitemShut
  {NoStop}%
\bibitem [{\citenamefont {Takeuchi}\ \emph {et~al.}(2009)\citenamefont
  {Takeuchi}, \citenamefont {Kuroda}, \citenamefont {Chaté},\ and\
  \citenamefont {Sano}}]{Takeuchi2009}%
  \BibitemOpen
  \bibfield  {author} {\bibinfo {author} {\bibfnamefont {K.~A.}\ \bibnamefont
  {Takeuchi}}, \bibinfo {author} {\bibfnamefont {M.}~\bibnamefont {Kuroda}},
  \bibinfo {author} {\bibfnamefont {H.}~\bibnamefont {Chaté}},\ and\ \bibinfo
  {author} {\bibfnamefont {M.}~\bibnamefont {Sano}},\ }\bibfield  {journal}
  {\bibinfo  {journal} {Phys. Rev. E}\ }\textbf {\bibinfo {volume} {80}},\
  \href {https://doi.org/10.1103/PhysRevE.80.051116}
  {10.1103/PhysRevE.80.051116} (\bibinfo {year} {2009})\BibitemShut {NoStop}%
\bibitem [{\citenamefont {Lemoult}\ \emph {et~al.}(2016)\citenamefont
  {Lemoult}, \citenamefont {Shi}, \citenamefont {Avila}, \citenamefont
  {Jalikop}, \citenamefont {Avila},\ and\ \citenamefont {Hof}}]{Lemoult2016}%
  \BibitemOpen
  \bibfield  {author} {\bibinfo {author} {\bibfnamefont {G.}~\bibnamefont
  {Lemoult}}, \bibinfo {author} {\bibfnamefont {L.}~\bibnamefont {Shi}},
  \bibinfo {author} {\bibfnamefont {K.}~\bibnamefont {Avila}}, \bibinfo
  {author} {\bibfnamefont {S.~V.}\ \bibnamefont {Jalikop}}, \bibinfo {author}
  {\bibfnamefont {M.}~\bibnamefont {Avila}},\ and\ \bibinfo {author}
  {\bibfnamefont {B.}~\bibnamefont {Hof}},\ }\href
  {https://doi.org/10.1038/nphys3675} {\bibfield  {journal} {\bibinfo
  {journal} {Nat. Phys.}\ }\textbf {\bibinfo {volume} {12}},\ \bibinfo {pages}
  {254} (\bibinfo {year} {2016})}\BibitemShut {NoStop}%
\bibitem [{\citenamefont {Barreales}\ \emph {et~al.}(2020)\citenamefont
  {Barreales}, \citenamefont {Mel{\'{e}}ndez}, \citenamefont {Cuerno},\ and\
  \citenamefont {Ruiz-Lorenzo}}]{Barreales2020}%
  \BibitemOpen
  \bibfield  {author} {\bibinfo {author} {\bibfnamefont {B.~G.}\ \bibnamefont
  {Barreales}}, \bibinfo {author} {\bibfnamefont {J.~J.}\ \bibnamefont
  {Mel{\'{e}}ndez}}, \bibinfo {author} {\bibfnamefont {R.}~\bibnamefont
  {Cuerno}},\ and\ \bibinfo {author} {\bibfnamefont {J.~J.}\ \bibnamefont
  {Ruiz-Lorenzo}},\ }\href {https://doi.org/10.1088/1742-5468/ab6a03}
  {\bibfield  {journal} {\bibinfo  {journal} {J. Stat. Mech.}\ ,\ \bibinfo
  {pages} {023203}} (\bibinfo {year} {2020})}\BibitemShut {NoStop}%
\bibitem [{\citenamefont {L{\'{o}}pez}\ \emph {et~al.}(1997)\citenamefont
  {L{\'{o}}pez}, \citenamefont {Rodr{\'{i}}guez},\ and\ \citenamefont
  {Cuerno}}]{Lopez1997}%
  \BibitemOpen
  \bibfield  {author} {\bibinfo {author} {\bibfnamefont {J.~M.}\ \bibnamefont
  {L{\'{o}}pez}}, \bibinfo {author} {\bibfnamefont {M.~A.}\ \bibnamefont
  {Rodr{\'{i}}guez}},\ and\ \bibinfo {author} {\bibfnamefont {R.}~\bibnamefont
  {Cuerno}},\ }\href {https://doi.org/10.1016/S0378-4371(97)00375-0} {\bibfield
   {journal} {\bibinfo  {journal} {Physica A (Amsterdam)}\ }\textbf {\bibinfo
  {volume} {246}},\ \bibinfo {pages} {329} (\bibinfo {year}
  {1997})}\BibitemShut {NoStop}%
\bibitem [{\citenamefont {Young}(2015)}]{Young2015}%
  \BibitemOpen
  \bibfield  {author} {\bibinfo {author} {\bibfnamefont {P.}~\bibnamefont
  {Young}},\ }\href {https://doi.org/10.1007/978-3-319-19051-8} {\emph
  {\bibinfo {title} {{Everything You Wanted to Know About Data Analysis and
  Fitting but Were Afraid to Ask}}}}\ (\bibinfo  {publisher} {Springer},\
  \bibinfo {address} {London},\ \bibinfo {year} {2015})\BibitemShut {NoStop}%
\bibitem [{\citenamefont {Efron}(1982)}]{Efron1982}%
  \BibitemOpen
  \bibfield  {author} {\bibinfo {author} {\bibfnamefont {B.}~\bibnamefont
  {Efron}},\ }\href {https://doi.org/10.1137/1.9781611970319} {\emph {\bibinfo
  {title} {{The jackknife, the bootstrap, and other resampling plans}}}}\
  (\bibinfo  {publisher} {Society for Industrial and Applied Mathematics},\
  \bibinfo {address} {Philadelphia},\ \bibinfo {year} {1982})\ p.~\bibinfo
  {pages} {92}\BibitemShut {NoStop}%
\bibitem [{\citenamefont {Siegert}(1996)}]{Siegert1996}%
  \BibitemOpen
  \bibfield  {author} {\bibinfo {author} {\bibfnamefont {M.}~\bibnamefont
  {Siegert}},\ }\href {https://doi.org/10.1103/PhysRevE.53.3209} {\bibfield
  {journal} {\bibinfo  {journal} {Phys. Rev. E}\ }\textbf {\bibinfo {volume}
  {53}},\ \bibinfo {pages} {3209} (\bibinfo {year} {1996})}\BibitemShut
  {NoStop}%
\bibitem [{\citenamefont {L\'opez}(1999)}]{Lopez1999}%
  \BibitemOpen
  \bibfield  {author} {\bibinfo {author} {\bibfnamefont {J.~M.}\ \bibnamefont
  {L\'opez}},\ }\href {https://doi.org/10.1103/PhysRevLett.83.4594} {\bibfield
  {journal} {\bibinfo  {journal} {Phys. Rev. Lett.}\ }\textbf {\bibinfo
  {volume} {83}},\ \bibinfo {pages} {4594} (\bibinfo {year}
  {1999})}\BibitemShut {NoStop}%
\bibitem [{\citenamefont {Barreales}\ \emph
  {et~al.}(2023{\natexlab{a}})\citenamefont {Barreales}, \citenamefont
  {Mel{\'{e}}ndez}, \citenamefont {Cuerno},\ and\ \citenamefont
  {Ruiz-Lorenzo}}]{datasetchi}%
  \BibitemOpen
  \bibfield  {author} {\bibinfo {author} {\bibfnamefont {B.~G.}\ \bibnamefont
  {Barreales}}, \bibinfo {author} {\bibfnamefont {J.~J.}\ \bibnamefont
  {Mel{\'{e}}ndez}}, \bibinfo {author} {\bibfnamefont {R.}~\bibnamefont
  {Cuerno}},\ and\ \bibinfo {author} {\bibfnamefont {J.~J.}\ \bibnamefont
  {Ruiz-Lorenzo}},\ }\href {https://doi.org/10.5281/zenodo.7794480}
  {10.5281/zenodo.7794480} (\bibinfo {year} {2023}{\natexlab{a}})\BibitemShut
  {NoStop}%
\bibitem [{\citenamefont {Monthus}\ and\ \citenamefont
  {Garel}(2008)}]{Monthus2008}%
  \BibitemOpen
  \bibfield  {author} {\bibinfo {author} {\bibfnamefont {C.}~\bibnamefont
  {Monthus}}\ and\ \bibinfo {author} {\bibfnamefont {T.}~\bibnamefont
  {Garel}},\ }\href {https://doi.org/10.1088/1742-5468/2008/01/P01008}
  {\bibfield  {journal} {\bibinfo  {journal} {J. Stat. Mech.}\ }\textbf
  {\bibinfo {volume} {2008}},\ \bibinfo {pages} {P01008} (\bibinfo {year}
  {2008})}\BibitemShut {NoStop}%
\bibitem [{\citenamefont {Majumdar}\ and\ \citenamefont
  {Schehr}(2014)}]{majumdar2014}%
  \BibitemOpen
  \bibfield  {author} {\bibinfo {author} {\bibfnamefont {S.~N.}\ \bibnamefont
  {Majumdar}}\ and\ \bibinfo {author} {\bibfnamefont {G.}~\bibnamefont
  {Schehr}},\ }\href {https://doi.org/10.1088/1742-5468/2014/01/P01012}
  {\bibfield  {journal} {\bibinfo  {journal} {J. Stat. Mech.}\ }\textbf
  {\bibinfo {volume} {2014}},\ \bibinfo {pages} {P01012} (\bibinfo {year}
  {2014})}\BibitemShut {NoStop}%
\bibitem [{\citenamefont {Kim}\ \emph {et~al.}(1991)\citenamefont {Kim},
  \citenamefont {Moore},\ and\ \citenamefont {Bray}}]{kim1991}%
  \BibitemOpen
  \bibfield  {author} {\bibinfo {author} {\bibfnamefont {J.~M.}\ \bibnamefont
  {Kim}}, \bibinfo {author} {\bibfnamefont {M.~A.}\ \bibnamefont {Moore}},\
  and\ \bibinfo {author} {\bibfnamefont {A.~J.}\ \bibnamefont {Bray}},\ }\href
  {https://doi.org/10.1103/PhysRevA.44.2345} {\bibfield  {journal} {\bibinfo
  {journal} {Phys. Rev. A}\ }\textbf {\bibinfo {volume} {44}},\ \bibinfo
  {pages} {2345} (\bibinfo {year} {1991})}\BibitemShut {NoStop}%
\bibitem [{\citenamefont {Barreales}\ \emph
  {et~al.}(2023{\natexlab{b}})\citenamefont {Barreales}, \citenamefont
  {Mel{\'{e}}ndez}, \citenamefont {Cuerno},\ and\ \citenamefont
  {Ruiz-Lorenzo}}]{datasetC1}%
  \BibitemOpen
  \bibfield  {author} {\bibinfo {author} {\bibfnamefont {B.~G.}\ \bibnamefont
  {Barreales}}, \bibinfo {author} {\bibfnamefont {J.~J.}\ \bibnamefont
  {Mel{\'{e}}ndez}}, \bibinfo {author} {\bibfnamefont {R.}~\bibnamefont
  {Cuerno}},\ and\ \bibinfo {author} {\bibfnamefont {J.~J.}\ \bibnamefont
  {Ruiz-Lorenzo}},\ }\href {https://doi.org/10.5281/zenodo.7788436}
  {10.5281/zenodo.7788436} (\bibinfo {year} {2023}{\natexlab{b}})\BibitemShut
  {NoStop}%
\bibitem [{\citenamefont {Bornemann}(2010)}]{bornemann2010}%
  \BibitemOpen
  \bibfield  {author} {\bibinfo {author} {\bibfnamefont {F.}~\bibnamefont
  {Bornemann}},\ }\href {https://doi.org/10.1090/S0025-5718-09-02280-7}
  {\bibfield  {journal} {\bibinfo  {journal} {Math. Comput.}\ }\textbf
  {\bibinfo {volume} {79}},\ \bibinfo {pages} {871} (\bibinfo {year}
  {2010})}\BibitemShut {NoStop}%
\bibitem [{\citenamefont {Ben-Avraham}\ \emph {et~al.}(1990)\citenamefont
  {Ben-Avraham}, \citenamefont {Burschka},\ and\ \citenamefont
  {Doering}}]{Ben-Avraham1990}%
  \BibitemOpen
  \bibfield  {author} {\bibinfo {author} {\bibfnamefont {D.}~\bibnamefont
  {Ben-Avraham}}, \bibinfo {author} {\bibfnamefont {M.~A.}\ \bibnamefont
  {Burschka}},\ and\ \bibinfo {author} {\bibfnamefont {C.~R.}\ \bibnamefont
  {Doering}},\ }\href {https://doi.org/10.1007/BF01025990} {\bibfield
  {journal} {\bibinfo  {journal} {J. Stat. Phys.}\ }\textbf {\bibinfo {volume}
  {60}},\ \bibinfo {pages} {695} (\bibinfo {year} {1990})}\BibitemShut
  {NoStop}%
\bibitem [{\citenamefont {Pechenik}\ and\ \citenamefont
  {Levine}(1999)}]{Pechenik1999}%
  \BibitemOpen
  \bibfield  {author} {\bibinfo {author} {\bibfnamefont {L.}~\bibnamefont
  {Pechenik}}\ and\ \bibinfo {author} {\bibfnamefont {H.}~\bibnamefont
  {Levine}},\ }\href {https://doi.org/10.1103/PhysRevE.59.3893} {\bibfield
  {journal} {\bibinfo  {journal} {Phys. Rev. E}\ }\textbf {\bibinfo {volume}
  {59}},\ \bibinfo {pages} {3893} (\bibinfo {year} {1999})}\BibitemShut
  {NoStop}%
\bibitem [{\citenamefont {Doering}\ \emph {et~al.}(2003)\citenamefont
  {Doering}, \citenamefont {Mueller},\ and\ \citenamefont
  {Smereka}}]{Doering2003}%
  \BibitemOpen
  \bibfield  {author} {\bibinfo {author} {\bibfnamefont {C.~R.}\ \bibnamefont
  {Doering}}, \bibinfo {author} {\bibfnamefont {C.}~\bibnamefont {Mueller}},\
  and\ \bibinfo {author} {\bibfnamefont {P.}~\bibnamefont {Smereka}},\ }\href
  {https://doi.org/10.1016/S0378-4371(03)00203-6} {\bibfield  {journal}
  {\bibinfo  {journal} {Phys. A (Amsterdam)}\ }\textbf {\bibinfo {volume}
  {325}},\ \bibinfo {pages} {243} (\bibinfo {year} {2003})}\BibitemShut
  {NoStop}%
\bibitem [{\citenamefont {Nesic}\ \emph {et~al.}(2014)\citenamefont {Nesic},
  \citenamefont {Cuerno},\ and\ \citenamefont {Moro}}]{Nesic2014}%
  \BibitemOpen
  \bibfield  {author} {\bibinfo {author} {\bibfnamefont {S.}~\bibnamefont
  {Nesic}}, \bibinfo {author} {\bibfnamefont {R.}~\bibnamefont {Cuerno}},\ and\
  \bibinfo {author} {\bibfnamefont {E.}~\bibnamefont {Moro}},\ }\href
  {https://doi.org/10.1103/PhysRevLett.113.180602} {\bibfield  {journal}
  {\bibinfo  {journal} {Phys. Rev. Lett.}\ }\textbf {\bibinfo {volume} {113}},\
  \bibinfo {pages} {180602} (\bibinfo {year} {2014})}\BibitemShut {NoStop}%
\bibitem [{\citenamefont {Song}\ and\ \citenamefont {Kim}(2006)}]{Song2006}%
  \BibitemOpen
  \bibfield  {author} {\bibinfo {author} {\bibfnamefont {H.~S.}\ \bibnamefont
  {Song}}\ and\ \bibinfo {author} {\bibfnamefont {J.~M.}\ \bibnamefont {Kim}},\
  }\href@noop {} {\bibfield  {journal} {\bibinfo  {journal} {J. Korean Phys.
  Soc.}\ }\textbf {\bibinfo {volume} {49}},\ \bibinfo {pages} {1520} (\bibinfo
  {year} {2006})}\BibitemShut {NoStop}%
\bibitem [{\citenamefont {Song}\ and\ \citenamefont {Kim}(2008)}]{Song2008}%
  \BibitemOpen
  \bibfield  {author} {\bibinfo {author} {\bibfnamefont {H.~S.}\ \bibnamefont
  {Song}}\ and\ \bibinfo {author} {\bibfnamefont {J.~M.}\ \bibnamefont {Kim}},\
  }\href {https://doi.org/10.3938/jkps.52.166} {\bibfield  {journal} {\bibinfo
  {journal} {J. Korean Phys. Soc.}\ }\textbf {\bibinfo {volume} {52}},\
  \bibinfo {pages} {166} (\bibinfo {year} {2008})}\BibitemShut {NoStop}%
\bibitem [{\citenamefont {Song}\ and\ \citenamefont {Kim}(2011)}]{Song2011}%
  \BibitemOpen
  \bibfield  {author} {\bibinfo {author} {\bibfnamefont {H.~S.}\ \bibnamefont
  {Song}}\ and\ \bibinfo {author} {\bibfnamefont {J.~M.}\ \bibnamefont {Kim}},\
  }\href {https://doi.org/10.1088/1742-5468/2011/09/P09021} {\bibfield
  {journal} {\bibinfo  {journal} {J. Stat. Mech.}\ ,\ \bibinfo {pages}
  {P09021}} (\bibinfo {year} {2011})}\BibitemShut {NoStop}%
\bibitem [{\citenamefont {Pikovsky}\ \emph {et~al.}(2001)\citenamefont
  {Pikovsky}, \citenamefont {Rosenblum},\ and\ \citenamefont
  {Kurths}}]{Pikovsky2001}%
  \BibitemOpen
  \bibfield  {author} {\bibinfo {author} {\bibfnamefont {A.}~\bibnamefont
  {Pikovsky}}, \bibinfo {author} {\bibfnamefont {M.}~\bibnamefont
  {Rosenblum}},\ and\ \bibinfo {author} {\bibfnamefont {J.}~\bibnamefont
  {Kurths}},\ }\href {https://doi.org/10.1017/CBO9780511755743} {\emph
  {\bibinfo {title} {Synchronization: A Universal Concept in Nonlinear
  Sciences}}}\ (\bibinfo  {publisher} {Cambridge University Press},\ \bibinfo
  {address} {Cambridge, UK},\ \bibinfo {year} {2001})\BibitemShut {NoStop}%
\bibitem [{\citenamefont {Arenas}\ \emph {et~al.}(2008)\citenamefont {Arenas},
  \citenamefont {Díaz-Guilera}, \citenamefont {Kurths}, \citenamefont
  {Moreno},\ and\ \citenamefont {Zhou}}]{Arenas2008}%
  \BibitemOpen
  \bibfield  {author} {\bibinfo {author} {\bibfnamefont {A.}~\bibnamefont
  {Arenas}}, \bibinfo {author} {\bibfnamefont {A.}~\bibnamefont
  {Díaz-Guilera}}, \bibinfo {author} {\bibfnamefont {J.}~\bibnamefont
  {Kurths}}, \bibinfo {author} {\bibfnamefont {Y.}~\bibnamefont {Moreno}},\
  and\ \bibinfo {author} {\bibfnamefont {C.}~\bibnamefont {Zhou}},\ }\href
  {https://doi.org/https://doi.org/10.1016/j.physrep.2008.09.002} {\bibfield
  {journal} {\bibinfo  {journal} {Phys. Rep.}\ }\textbf {\bibinfo {volume}
  {469}},\ \bibinfo {pages} {93} (\bibinfo {year} {2008})}\BibitemShut
  {NoStop}%
\end{thebibliography}
\end{document}